\documentclass[11pt]{article}
\usepackage{epsfig} 
\usepackage{amssymb}
\usepackage{graphics}
\newcommand{\sect}[1]{ \section{#1} \setcounter{equation}{0} } 

\newcommand{\half}{\mbox{\small{$\frac{1}{2}$}}} 
\newcommand{\third}{\mbox{\small{$\frac{1}{3}$}}} 
 
\newcommand{\pgone}{\psi^{(1)}({\third})}
\newcommand{\pgthree}{\psi^{(3)}({\third})}
\newcommand{\pgfive}{\psi^{(5)}({\third})}
\newcommand{\Nc}{N_{\!c}}
\newcommand{\Nf}{N_{\!f}}

\newcommand{\NA}{N_{\!A}}
\newcommand{\bare}{\mbox{\footnotesize{o}}}

\newcommand{\pslash}{p \! \! \! /}
\newcommand{\Dslash}{D \! \! \! \! /}

\newcommand{\QEDss}{\mbox{\scriptsize{QED}}}
\newcommand{\MSbar}{\overline{\mbox{MS}}}

\newcommand{\MSbarss}{\overline{\mbox{\scriptsize{MS}}}}

\newcommand{\MOMts}{\widetilde{\mbox{\footnotesize{MOM}}}}

\newcommand{\MOMtccgzcs}{\widetilde{\mbox{\footnotesize{MOM}}}_{ccg0c}}
\newcommand{\MOMtccgzcss}{\widetilde{\mbox{\scriptsize{MOM}}}_{ccg0c}}

\newcommand{\MOMtccgzgs}{\widetilde{\mbox{\footnotesize{MOM}}}_{ccg0g}}
\newcommand{\MOMtccgzgss}{\widetilde{\mbox{\scriptsize{MOM}}}_{ccg0g}}

\newcommand{\MOMtgggzgs}{\widetilde{\mbox{\footnotesize{MOM}}}_{ggg0g}}
\newcommand{\MOMtgggzgss}{\widetilde{\mbox{\scriptsize{MOM}}}_{ggg0g}}

\newcommand{\MOMtgggzggs}{\widetilde{\mbox{\footnotesize{MOM}}}_{ggg0gg}}
\newcommand{\MOMtgggzggss}{\widetilde{\mbox{\scriptsize{MOM}}}_{ggg0gg}}

\newcommand{\MOMtqqgzgs}{\widetilde{\mbox{\footnotesize{MOM}}}_{qqg0g}}
\newcommand{\MOMtqqgzgss}{\widetilde{\mbox{\scriptsize{MOM}}}_{qqg0g}}

\newcommand{\MOMtqqgzgTs}{\widetilde{\mbox{\footnotesize{MOM}}}_{qqg0gT}}
\newcommand{\MOMtqqgzgTss}{\widetilde{\mbox{\scriptsize{MOM}}}_{qqg0gT}}

\newcommand{\MOMtqqgzqs}{\widetilde{\mbox{\footnotesize{MOM}}}_{qqg0q}}
\newcommand{\MOMtqqgzqss}{\widetilde{\mbox{\scriptsize{MOM}}}_{qqg0q}}

\newcommand{\MOMtqqgzqTs}{\widetilde{\mbox{\footnotesize{MOM}}}_{qqg0qT}}
\newcommand{\MOMtqqgzqTss}{\widetilde{\mbox{\scriptsize{MOM}}}_{qqg0qT}}

\newcommand{\MOMtstars}{\widetilde{\mbox{\footnotesize{MOM}}}_{\ast}}
\newcommand{\MOMtstarss}{\widetilde{\mbox{\scriptsize{MOM}}}_{\ast}}
\newcommand{\MOMc}{\mbox{MOMc}}

\newcommand{\MOMg}{\mbox{MOMg}}

\newcommand{\MOMq}{\mbox{MOMq}}

\newcommand{\MOMqss}{\mbox{\scriptsize{MOMq}}}
\newcommand{\mMOM}{\mbox{mMOM}}

\newcommand{\mMOMss}{\mbox{\scriptsize{mMOM}}}
\newcommand{\RI}{\mbox{RI${}^\prime$}}
\newcommand{\RIs}{\mbox{\footnotesize{RI${}^\prime$}}}

\newcommand{\MOMteepzpTs}{\widetilde{\mbox{\footnotesize{MOM}}}_{eep0pT}}
\newcommand{\MOMteepzpTss}{\widetilde{\mbox{\scriptsize{MOM}}}_{eep0pT}}

\begin{document}

\title{Explicit no-$\pi^2$ renormalization schemes in QCD at five loops}

\author{J.A. Gracey, \\ Theoretical Physics Division, \\ 
Department of Mathematical Sciences, \\ University of Liverpool, \\ P.O. Box 
147, \\ Liverpool, \\ L69 3BX, \\ United Kingdom.} 

%\date{November, 2023.}
\date{}

\maketitle 

\vspace{5cm} 
\noindent 
{\bf Abstract.} We examine a variety of renormalization schemes in QCD based on
its $3$-point vertices where the $\beta$-functions, gluon, ghost, quark and 
quark mass anomalous dimensions in each scheme do not depend on $\zeta_4$ or
$\zeta_6$ in an arbitrary linear covariant gauge at five loops. We comment on
the $C$-scheme.

%\vspace{-16.2cm}
\vspace{-15.0cm}
\hspace{13.4cm}
{\bf LTH 1357}

\newpage 

\sect{Introduction.}

Over the last decade or so remarkable advances have been made in computing the 
renormalization group functions of four dimensional quantum field theories to 
very high orders. Perhaps the most significant of these is the advancement of 
Quantum Chromodynamics (QCD) to five loops,
\cite{1,2,3,4,5,6,7,8,9,10,11,12,13,14,15,16,17,18,19,20,21,22}. This has been 
made possible, for instance, due to the development of the {\sc Forcer} 
package, \cite{23,24}, written in the symbolic manipulation language 
{\sc Form}, \cite{25,26}. The package evaluates four loop massless Feynman 
integrals contributing to $2$-point functions in $d$-dimensions as well as 
carrying out the expansion in $\epsilon$ where $d$~$=$~$4$~$-$~$2\epsilon$. Its
application to five loop computations has been possible through the use of the 
$R^\star$ operation \cite{27,28,29,30,31,32,33,34,35,36} and the four loop 
{\sc Forcer} package itself. While gauge theories are central to the Standard 
Model, progress in renormalizing scalar theories has advanced to an even higher 
loop order, \cite{37,38}. Such results are important too as they give insight 
into the number basis structure of the renormalization group functions that 
ought to have parallels in gauge theories when they are advanced to the next 
level. For instance, it is widely known that in the modified minimal 
subtraction ($\MSbar$) scheme, \cite{39,40}, the renormalization group 
functions involve the Riemann $\zeta$-function up to $\zeta_{11}$ at seven 
loops in the $\phi^4$ $\beta$-function, \cite{38}. In addition the multiple 
zetas $\zeta_{5,3}$ and $\zeta_{5,3,3}$ first appear at six and seven loops 
respectively, \cite{37,38}. On top of this a new period, denoted by $P_{7,11}$ 
in \cite{38}, occurs which is believed to be inexpressible in terms of multiple
zetas although it can be written in terms of multiple polylogarithms of the 
sixth roots of unity, \cite{38}. Beyond seven loops it is conjectured that 
other new periods will arise, \cite{38}. 

Knowledge of such potential structures is important in devising efficient 
computational tools to carry out future higher order renormalization. From the 
high loop order data that has been accumulated over many years several features
have become apparent. For instance, at $L$ loops no $\zeta_n$ can be present 
where $n$~$>$~$(2L-3)$ beginning at $L$~$=$~$3$. Earlier numbers in the 
$\zeta_n$ sequence, the Euler-Mascheroni constant $\gamma$ and $\zeta_2$, are
absent. The former due to the choice of the $\MSbar$ scheme, \cite{39,40}, over
its predecessor the minimal subtraction scheme, and the latter as it actually
cancels when the renormalization group functions are compiled from the
renormalization constants. However we qualify this $\zeta_n$ structure by
noting that it is primarily based on what occurs in a scalar theory. When
symmetries are present the structure may be simpler. This is the case in the
QCD $\beta$-function as $\zeta_3$ first appears at four loops and $\zeta_5$
at five loops in the $\MSbar$ scheme, \cite{11,16,18,19,21,22}. This is not the
situation for the remaining core renormalization group functions which follow 
the $\zeta_{2L-3}$ pattern for their first occurence. By core we will mean
throughout the gluon, Faddeev-Popov ghost, quark and quark mass anomalous 
dimensions. The absence of $\zeta_3$ at three loops in the $\beta$-function is 
due to gauge symmetry and the Slavnov-Taylor identity, \cite{41,42}. In schemes
other than $\MSbar$, such as the minimal momentum ($\mMOM$) scheme of 
\cite{43}, the $\zeta_{2L-3}$ pattern appears for the first time at $L$ loops 
in the $\beta$-function for $L$~$\geq$~$3$.

Another property of the number structure of the QCD $\beta$-function that is
not unrelated to that for the odd zetas concerns the location of the even ones.
Ordinarily $\zeta_{2L-4}$ would be expected to first appear at $L$ loops for 
$L$~$\geq$~$4$ but in the $\MSbar$ scheme this happens at $(L+1)$ loops. We
recall that $\zeta_{2n}$ is proportional to $\pi^{2n}$ for $n$~$\geq$~$1$. As 
before this even zeta structure is not mirrored in the other core 
renormalization group functions. Over the years the location or otherwise of 
the sequence of even zetas in renormalization group functions in various 
theories has been the subject of more than passing interest. Indeed a debate 
has ensued as to whether or not there is a natural way that 
$\zeta_4$~$=$~$\frac{\pi^4}{90}$ and $\zeta_6$~$=$~$\frac{\pi^6}{945}$ can be 
excluded and in what circumstances. One aspect of what is meant by this is 
whether there is an appropriate renormalization prescription that produces this
scenario at a fundamental level. Another line of study is to examine whether 
there are so-called redefinitions of the odd zeta sequences that are universal 
across theories and schemes. Perhaps the pivotal instance where this was 
illuminated was provided in \cite{44}. The focus there was on the perturbative 
structure of the Adler $D$-function in the $\MSbar$ scheme and the absence of 
$\zeta_4$ was noted in the $O(a^4)$ term corresponding to the evaluation of 
five loop graphs, a property which is also shared with the R ratio at the same 
order. Prior to \cite{44} it had been shown that $\zeta_4$ appeared in the five
loop quark mass anomalous dimension, \cite{20}, as well as at four loops, 
\cite{12,13}. A detailed analysis of these observations in \cite{44} suggested 
that for the renormalization group functions there was a systematic way of 
removing the even zetas by using an $\epsilon$ dependent transformation based 
on a connection with their odd partners. This was robustly examined for other 
QCD quantities as well as in other theories, \cite{44}. Moreover the 
observation was grounded from a renormalization group perspective. The idea was
to concentrate on the location of the $\zeta_n$ sequence within the $\epsilon$ 
expansion of contributions to Feynman integrals that will affect the higher 
loop order counterterms. Aspects of this perspective were verified at very high
orders in $\epsilon$ at several orders in the large $\Nf$ expansion, 
\cite{44,45}, where $\Nf$ is the number of quarks. More recent developments 
have followed in the multicoupling context and for theories with supersymmetry 
for instance, \cite{46}.

In field theory language the ideas of \cite{44,45,46} should translate to 
changing to a different renormalization scheme. This is borne out by a second 
line of investigation which is to find a renormalization scheme that 
automatically produces a $\zeta_{2n}$ free set of renormalization group 
functions. There are already several pointers to such a scheme but in each case
it appears that the picture is not complete. For instance, in \cite{47} three 
schemes were introduced in QCD, and allocated the scheme label $\MOMts$, with 
the low orders of the $\beta$-function determined in a specific gauge. The 
three schemes were based on each of the three $3$-point vertices. The 
perturbative structure of these $\beta$-functions was extended to four loops in
the Landau gauge in \cite{48,49} as well as an additional scheme based on a 
different renormalization condition. That extra scheme had been introduced in 
\cite{50} and the $\beta$-function constructed to three loops before being 
extended to four loops in \cite{49}. The $\MOMts$ scheme setup of \cite{47,49} 
is a variant of the symmetric point momentum subtraction (MOM) ones of 
\cite{51,52}. There each $3$-point vertex was evaluated to the finite part and 
the finite part absorbed into the coupling renormalization constant with the 
same subtraction prescription used for the $2$-point functions. In the $\MOMts$
schemes of \cite{47,49} the same subtraction is carried out but the $3$-point 
vertices are first evaluated where one of the external momenta is set to zero. 
In other words in a situation where there is only one external momentum rather 
than two independent momenta as in the case of the MOM schemes of \cite{51,52}. 
Throughout we will use the notation that $\MOMts$ is the umbrella term for the 
momentum configuration used in \cite{47,49} for $3$-point vertices and use that
term when referring to articles where the prescription was in fact employed but
a different label, such as MOM, was given. This is to avoid confusion since MOM
is more commonly used for the schemes of \cite{51,52}. While \cite{49} provided
a four loop analysis the absence of $\zeta_4$ was obvious though not surprising
given it does not occur in the $\MSbar$ scheme at that order. Where the absence
of $\zeta_4$ became more significant was in the construction of the five loop 
$\beta$-function in the $\MOMts$ scheme in Quantum Electrodynamics (QED), 
\cite{53}. There the Ward-Takahashi identity was exploited to deduce the 
$\beta$-function from the photon renormalization constant with the finite part 
of the $2$-point function removed at the subtraction point. The resulting 
$\beta$-function was devoid of $\zeta_4$ and $\zeta_6$. This example has since 
been classified as occuring in an anomalous dimension (AD) theory, \cite{46}. 
Such a theory is one where the $\beta$-function is deduced via a symmetry, 
which could be gauge symmetry for example, as it is directly related to the 
anomalous dimension of one or a combination of fields. 

Another similar AD example is the Wess-Zumino theory, \cite{54}, which is a 
four dimensional supersymmetric model. Its coupling constant renormalization is
not independent since it is related to the field anomalous dimension via a 
supersymmetry Ward identity. In \cite{55} the three loop $\MOMts$ 
$\beta$-function was determined and subsequently extended to five loops in 
\cite{56}. The four loop $\MSbar$ $\beta$-function was provided in \cite{57}
which corrected errors in earlier lower loop calculations. What was observed in
the five loop $\MOMts$ $\beta$-function, \cite{56}, was the absence of $\zeta_4$
and $\zeta_6$ similar to \cite{53}. Moreover the sector of the $\MOMts$ 
$\beta$-function that corresponded to the iteration of the one loop bubble 
agreed precisely with the Hopf algebra $\MOMts$ construction of that set of 
graphs presented in \cite{58}. In fact iterating the Hopf algebra result well
beyond the five loop order of \cite{56} showed that there were no $\zeta_{2n}$ 
contributions to the very high order that was recorded in \cite{58}. While such
one loop bubble contributions are not the complete set of graphs it does 
provide strong evidence in a concrete example that the absence of even zetas 
may be a more fundamental property of the $\MOMts$ scheme at least in the case 
of AD theories. One reason why the $\MOMts$ scheme would offer a more 
satisfactory way to proceed, aside from being based on a Lagrangian and 
systematically implemented by a renormalization prescription, is that it is not
clear what effect the detailed examination of the even zeta cancellation of 
\cite{44,45} has on the remaining non-even zeta part of renormalization group 
functions. Ultimately it is the full renormalization group functions that are 
necessary for any application involving observables. Some progress in that 
direction has been provided in the $C$-scheme of \cite{59,60}. This is a scheme
that depends on a parameter $C$ which is used as a measure of the scheme 
dependence of the coupling constant. One aspect of it is the claim that 
$\zeta_4$ terms are not present in the $C$-scheme versions of the Adler 
$D$-function and several operator correlation functions including that of the 
field strength, \cite{60}. Moreover the mapping of the $\MSbar$ coupling 
constant to its $C$-scheme counterpart was discussed in depth in an analysis of
the $\zeta_4$ and $\zeta_6$ dependence of the four loop anomalous dimensions of
flavour non-singlet and singlet twist-$2$ operators central to deep inelastic 
scattering for several low moments, \cite{61}. A variation of the $C$-scheme 
theme was explored in the $\hat{G}$-scheme provided in \cite{44}. Another 
approach to the absence of $\pi^2$ contributions in correlation functions was 
examined in \cite{62} through the application of the Landau-Khalatnikov-Fradkin
transformation.

In other words based on the evidence discussed so far the finite subtraction 
approach is the most promising to pursue. However in compiling this overview 
for the $\MOMts$ scheme and the zeta mapping analyses of \cite{44,45,46} what 
appears to be absent for the former is a full renormalization of each 
Lagrangian and in particular that of QCD for a general colour group and 
arbitrary linear covariant gauge. The main focus generally has been on the 
$\beta$-function compared with occasional interest in the core anomalous 
dimensions for non-AD theories or additionally in the case of a gauge theory 
only one specific gauge, the Landau one, has been considered, \cite{49}. The 
most recent $\MOMts$ schemes recorded for QCD were at four loops, \cite{49}, 
although the five loop $\MSbar$ renormalization group functions are all now
available for an arbitrary linear covariant gauge, 
\cite{1,2,3,4,5,6,7,8,9,10,11,12,13,14,15,16,17,18,19,20,21,22}. More recently
the determination of the five loop $\mMOM$ scheme core renormalization group 
functions has been completed, \cite{63,64}, for an arbitrary linear gauge. 
From the currently available $\MOMts$ expressions it has been indicated either 
explicitly or implicitly that there are no $\pi^2$ terms in the recorded 
$\beta$-functions. It is worth noting one case where the $\pi^2$ absence was
indeed highlighted which was in the determination of the five loop QED $\MOMts$ 
$\beta$-function, \cite{53}. Another reason to consider the core anomalous 
dimensions in a non-abelian gauge theory is the fact that the first location of
$\zeta_{2n}$ is different from the $\beta$-function. Therefore to be credible 
any $\MOMts$ style prescription has to produce a universal $\pi^2$ absence 
across all renormalization group functions as well as all gauges and colour 
groups for QCD. It is not clear how that would be effected at the level of the 
$\epsilon$ expansion of the $\zeta_n$ sequence within a suite of Feynman 
graphs. Having reviewed the background it is therefore the purpose of this 
article to balance both points of attack to understand the absence of $\pi^2$ 
in certain scenarios in non-abelian gauge theories. To achieve this we will 
provide various renormalization schemes in QCD based on the three $3$-point 
vertices of the linear covariant gauge fixed Lagrangian which extends previous 
work. This will be several more than those discussed in \cite{47,49}. In each 
of these schemes we will demonstrate that none of the renormalization group 
functions depend on $\zeta_4$ or $\zeta_6$ to {\em five} loops. One aim is to 
present as full and complete an analysis as possible using all available data 
and techniques especially properties of the renormalization group equation. 
This has the added benefit that the results we present can be used in future to 
examine the effects using such explicit no-$\pi^2$ schemes\footnote{We will 
refer to schemes that have no $\zeta_{2n}$ dependence as no-$\pi^2$ ones in 
order to avoid confusion with what is termed the no-$\pi$ theorem of 
\cite{44}.} have on phenomenological predictions along similar lines to those 
of \cite{59,60}.

The article is organized as follows. We review the basic methods and techniques
used for the analysis in Section $2$ and en route define our notation, 
conventions and the schemes we will focus on. The way we constructed the 
relevant renormalization constants is touched upon too. Our results are 
summarized and presented in Section $3$ as well as several checks. Having 
established our main goal we devote Section $4$ to a comparison of the QCD 
$\MOMts$ schemes with the $C$-scheme. Subsequently in Section $5$ we examine 
renormalization schemes in a larger context where other more general schemes 
are proposed and discussed. While the practical calculational study of such 
schemes is not as well advanced in terms of loop order, partly due to the 
absence of master integrals for $n$-point functions higher than three and 
general kinematics in analytic form, it is worth lighting the path ahead for 
future studies. Moreover once a deeper knowledge of the mathematical properties
of such master integrals is known more insight would be available to understand
the $\zeta_{2n}$ absence in the set of $\MOMts$ schemes at the level considered 
here and beyond. We summarize our findings in Section $6$. Finally two 
appendices follow with the first giving the Landau gauge $SU(3)$ five loop 
$\beta$-functions for the QCD $\MOMts$ schemes. The second appendix gives the 
$SU(3)$ Landau gauge anomalous dimensions for the scheme that closely connects 
with the $\mMOM$ scheme.

\sect{Formalism.}

We devote this section to recalling the relevant formalism of the 
renormalization group that we exploit to determine the five loop 
renormalization group functions as well as the method of computation to extract
the four loop renormalization constants. First as a reference point for our 
conventions we recall the bare QCD Lagrangian is 
\begin{equation}
L ~=~ -~ \frac{1}{4} G_{\bare \, \mu\nu}^a G_{\bare}^{a \, \mu\nu} ~-~
\frac{1}{2\alpha_{\bare}} (\partial^\mu A^a_{\bare \, \mu} )^2 ~-~
\bar{c}_{\bare}^a \left( \partial_\mu D_{\bare}^\mu c_{\bare} \right)^a ~+~
i \bar{\psi}_{\bare}^{iI} \left( \Dslash_{\bare} \psi_{\bare} \right)^{iI}
\label{lag}
\end{equation}
with ${}_{\bare}$ denoting a bare object where the respective gluon, ghost and
quark fields are $A^a_\mu$, $c^a$ and $\psi^{iI}$. We assume the fields lie in
a general Lie group and the indices take the ranges 
$1$~$\leq$~$a$~$\leq$~$\NA$, $1$~$\leq$~$i$~$\leq$~$\Nf$ and
$1$~$\leq$~$I$~$\leq$~$\Nc$ where $\NA$ is the adjoint representation dimension
and $\Nc$ is the fundamental representation dimension. As in \cite{63} we use 
the canonical linear covariant gauge fixing term with parameter $\alpha$ where 
$\alpha$~$=$~$0$ is the Landau gauge. The mapping of bare variables to their 
renormalized counterparts is defined by
\begin{equation}
A^{a \, \mu}_{\mbox{\footnotesize{o}}} ~=~ \sqrt{Z_A} \, A^{a \, \mu} ~~,~~
c^a_{\mbox{\footnotesize{o}}} ~=~ \sqrt{Z_c} \, c^a ~~,~~
\psi_{\mbox{\footnotesize{o}}} ~=~ \sqrt{Z_\psi} \psi ~~,~~
g_{\mbox{\footnotesize{o}}} ~=~ \mu^\epsilon Z_g \, g ~~,~~
\alpha_{\mbox{\footnotesize{o}}} ~=~ Z^{-1}_\alpha Z_A \, \alpha
\label{zdef}
\end{equation}
and we have introduced the mass scale $\mu$ that arises when the Lagrangian is
dimensionally regularized in $d$~$=$~$4$~$-$~$2\epsilon$ dimensions, which we
have used throughout, to ensure the coupling constant $g$ remains 
dimensionless. Once the renormalization constants are determined to a 
particular loop order they are encoded in the renormalization functions given 
by
\begin{equation}
\gamma_\phi(a,\alpha) ~=~ \mu \frac{\partial~}{\partial\mu} \ln Z_\phi ~~~,~~~
\beta(a,\alpha) ~=~ \mu \frac{\partial a}{\partial \mu} ~~~,~~~ 
\gamma_\alpha(a,\alpha) ~=~ \frac{\mu}{\alpha} 
\frac{\partial \alpha}{\partial \mu}
\label{rgedef}
\end{equation}
where $\phi$ is an element in the labelling set $\{A,c,\psi,m\}$ denoting the 
gluon, ghost, quark and quark mass renormalization respectively and
\begin{equation}
\mu \frac{\partial~}{\partial\mu} ~=~ \beta(a,\alpha) 
\frac{\partial~}{\partial a} ~+~ \alpha \gamma_\alpha(a,\alpha) 
\frac{\partial~}{\partial \alpha} 
\end{equation}
with $\alpha$ dependence included in the $\beta$-function. While the $\MSbar$ 
scheme $\beta$-function and quark mass anomalous dimensions are $\alpha$ 
independent, \cite{39}, this is not the case in general in schemes where a 
finite part is absorbed into the renormalization constant of the coupling. For 
completeness we note that (\ref{zdef}) and (\ref{rgedef}) lead to
\begin{eqnarray}
\gamma_A(a,\alpha) &=& \beta(a,\alpha) \frac{\partial}{\partial a} \ln Z_A ~+~ 
\alpha \gamma_\alpha(a,\alpha) \frac{\partial}{\partial \alpha} \ln Z_A 
\nonumber \\
\gamma_\alpha(a,\alpha) &=& \left[ \beta(a,\alpha) \frac{\partial}{\partial a}
\ln Z_\alpha ~-~ \gamma_A(a,\alpha) \right] \left[ 1 ~-~ \alpha
\frac{\partial}{\partial \alpha} \ln Z_\alpha \right]^{-1} \nonumber \\
\gamma_c(a,\alpha) &=& \beta(a,\alpha) \frac{\partial}{\partial a} \ln Z_c ~+~ 
\alpha \gamma_\alpha(a,\alpha) \frac{\partial}{\partial \alpha} \ln Z_c 
\nonumber \\
\gamma_\psi(a,\alpha) &=& \beta(a,\alpha) \frac{\partial}{\partial a} 
\ln Z_\psi ~+~ \alpha \gamma_\alpha(a,\alpha) \frac{\partial}{\partial \alpha} 
\ln Z_\psi 
\label{gammadef}
\end{eqnarray}
where $a$~$=$~$g^2/(16\pi^2)$. The relation between $Z_\alpha$ and
$\gamma_\alpha(a,\alpha)$ is more general than one would expect in the
canonical linear covariant gauge fixing. It is only when calculations in this
gauge determine $Z_\alpha$ to be unity that the more widely known relation
\begin{equation}
\gamma_\alpha(a,\alpha) ~=~ -~ \gamma_A(a,\alpha) 
\end{equation}
results which will be the case for each of the new schemes introduced here. 

As the renormalization of the parameters $a$ and $\alpha$ will be carried out
in several schemes one has to be able to map their running with $\mu$ from one
scheme to another. This is achieved by realizing the bare coupling parameter 
can be expressed in terms of $g$ in two different schemes producing 
\begin{equation}
g_{{\cal S}}(\mu) ~=~ \frac{Z_g^{\MSbarss}}{Z_g^{{\cal S}}} \, 
g_{\MSbarss}(\mu) ~~~,~~~
\alpha_{{\cal S}}(\mu) ~=~ \frac{Z_A^{\MSbarss}Z_\alpha^{{\cal S}}}
{Z_A^{{\cal S}}Z_\alpha^{\MSbarss}} \, \alpha_{\MSbarss}(\mu) 
\label{aalmap}
\end{equation}
where the relation for $\alpha$ follows from similar reasoning for 
$\alpha_{\bare}$. Throughout we will use ${\cal S}$ to indicate a scheme in
general. For convenience it is simpler to use the $\MSbar$ scheme as the base
or reference scheme for the discussion on the mapping of variables between
schemes. For the fields and quark mass we can construct similar conversion
functions which are given by
\begin{equation}
C^{{\cal S}}_\phi(a,\alpha) ~=~ 
\frac{Z_\phi^{{\cal S}}}{Z_\phi^{\MSbarss}} ~~~,~~~
C^{{\cal S}}_\alpha(a,\alpha) ~=~ \frac{Z_\alpha^{{\cal S}}Z_A^{\MSbarss}}
{Z_\alpha^{\MSbarss}Z_A^{{\cal S}}}
\end{equation}
where the latter is included for completeness. In each of these definitions one
has also to map the ${\cal S}$ scheme parameters to the $\MSbar$ ones by using,
in this instance,
\begin{eqnarray}
Z_\phi^{{\cal S}} ~=~ Z_\phi^{{\cal S}}
\left( a_{\cal S}(a,\alpha) , \alpha_{\cal S} (a,\alpha) \right) ~~~,~~~ 
Z_\alpha^{{\cal S}} ~=~ Z_\alpha^{{\cal S}}
\left( a_{\cal S}(a,\alpha) , \alpha_{\cal S} (a,\alpha) \right) 
\end{eqnarray}
as otherwise expressions with poles in $\epsilon$ will be present. Once the
parameter mappings and conversion functions have been explicitly established
the renormalization group equations can be deduced via
\begin{eqnarray}
\beta^{\cal S} ( a_{\cal S}, \alpha_{\cal S} ) &=&
\left[ \beta^{\mbox{$\MSbarss$}}( a_{\mbox{$\MSbarss$}} )
\frac{\partial a_{\cal S}}{\partial a_{\mbox{$\MSbarss$}}} \,+\,
\alpha_{\mbox{$\MSbarss$}} \gamma^{\mbox{$\MSbarss$}}_\alpha
( a_{\mbox{$\MSbarss$}}, \alpha_{\mbox{\footnotesize{$\MSbarss$}}} )
\frac{\partial a_{\cal S}}{\partial \alpha_{\mbox{$\MSbarss$}}}
\right]_{ \MSbarss \rightarrow {\cal S} } \nonumber \\
\gamma_\phi^{{\cal S}} ( a_{{\cal S}}, \alpha_{{\cal S}} )
&=& \!\! \! \left[ \gamma_\phi^{\MSbarss} \left(a_{\MSbarss}\right)
+ \beta^{\MSbarss}\left(a_{\MSbarss}\right)
\frac{\partial ~}{\partial a_{\MSbarss}} \ln C_\phi^{{\cal S}}
\left(a_{\MSbarss},\alpha_{\MSbarss}\right) \right. \nonumber \\
&& \left. +~ \alpha_{\MSbarss} \gamma^{\MSbarss}_\alpha
\left(a_{\MSbarss},\alpha_{\MSbarss}\right)
\frac{\partial ~}{\partial \alpha_{\MSbarss}}
\ln C_\phi^{{\cal S}} \left(a_{\MSbarss},\alpha_{\MSbarss}\right)
\right]_{ \MSbarss \rightarrow {\cal S} } ~.
\label{confunmap}
\end{eqnarray}
As the expressions on the left hand side are dependent on the scheme ${\cal S}$
variables the final stage in their construction is to use the inverse relations
to map $a_{\MSbarss}$ and $\alpha_{\MSbarss}$ to $a_{\cal S}$ and 
$\alpha_{\cal S}$ which is indicated by the restriction on each of the final
brackets. One useful aspect of this formalism is that if the renormalization
group functions are available in one scheme at $(L+1)$ loops then those of the
other scheme need only to be explicitly evaluated at $L$ loops in order to use
the conversion functions to find the parallel $(L+1)$ loop expressions. This
situation arises here since we will renormalize QCD in the set of $\MOMtstars$
schemes at four loops from which the $O(a^4)$ conversion functions can be
deduced. We use $\MOMtstars$ to denote the QCD $\MOMts$ schemes to be 
introduced shortly. Given that the five loop $\MSbar$ renormalization group 
functions are known, \cite{15,16,17,18,19,20,21,22}, then (\ref{confunmap}) can
be applied to determine the five loop $\MOMtstars$ renormalization group 
functions. This follows simply from (\ref{confunmap}) given that 
$\beta^{\MSbarss}\left(a_{\MSbarss}\right)$ and
$\gamma^{\MSbarss}_\alpha \left(a_{\MSbarss},\alpha_{\MSbarss}\right)$ are
$O(a^2)$ and $O(a)$ respectively. In this context it is worth remarking that
the renormalization group functions are an encoding or representation of the
renormalization constants computed explicitly from the field theory.

This process can of course be reversed to a certain extent. For instance it is 
well-known that in the $\MSbar$ scheme given the renormalization group 
functions at $L$ loops one can deduce the $L$ loop renormalization contants
correctly by integrating (\ref{gammadef}). However this needs to be qualified 
for schemes where the renormalization prescription involves the subtraction of 
a finite part of a Green's function such as the $\MOMtstars$ schemes considered 
here. This is because the final stage of extracting renormalization group 
functions from the renormalization constants is to lift the regularization. In 
dimensional regularization this would correspond to the limit 
$\epsilon$~$\to$~$0$. For the $\MSbar$ scheme the only $\epsilon$ dependence is
in the $O(a)$ term of the $\beta$-function; there is no $\epsilon$ dependence 
in any other core $\MSbar$ renormalization group functions. By contrast for 
schemes where there is a finite part in the subtraction to determine the 
renormalization constants, such as $\MOMtstars$, the coefficient of $a$ in each
term of the core renormalization group functions is linear in $\epsilon$ prior 
to removing the regularization. This $\epsilon$-dependent coefficient 
corresponds directly to the finite part of the renormalization constant and the 
dependence is rarely recorded in articles. Therefore in integrating 
(\ref{gammadef}) to find the $\MOMtstars$ renormalization constants, for 
example, one would have to initiate the process using the $\epsilon$-dependent 
renormalization group functions. 

Having outlined the relevant formalism to construct the renormalization group 
functions of our $\MOMtstars$ schemes we now focus on details. First there are 
three $3$-point vertices which in \cite{51,52} led to three distinct symmetric 
point schemes denoted by $\MOMg$, $\MOMc$ and $\MOMq$ based respectively on the
triple gluon, ghost-gluon and quark-gluon vertices. In these symmetric point 
schemes the renormalization prescription is to remove the poles as well as the 
finite parts at the momentum configuration where the squared momenta of the 
three external legs of the $3$-point vertices are equal. None of the underlying
external momenta are nullified. Equally the $2$-point functions are rendered 
finite by removing the finite part in addition to the poles. By contrast the 
$\MOMtstars$ schemes are defined from the $3$-point vertices but where the 
momentum configuration has one external momentum nullified at the outset. 
Similar to the schemes of \cite{51,52} the finite parts of both the $3$-point 
vertices for this configuration and the $2$-point functions are absorbed into 
the respective renormalization constants. One difference is that there are more
$\MOMtstars$ schemes, \cite{49}, than the three MOM schemes of \cite{51,52}. In
the first instance this is because for each of the ghost-gluon and quark-gluon 
vertices there are two possible external leg nullifications. These are either 
the gluon leg or one of the respective ghost or quark legs. One might suspect 
there is a third ghost-gluon vertex scheme given the asymmetric nature of the 
ghost-gluon interaction in (\ref{lag}). However it is trivial to see that the 
nullification of the $\bar{c}^a$ leg produces zero for each graph of the
$3$-point vertex in the linear covariant gauge. This is not the case for 
nonlinear covariant gauges for instance. For clarity it is worth recalling the 
tensor structure of the vertex functions for reference and to assist with the 
scheme definitions. Based on \cite{18,47} we have
\begin{eqnarray}
\left\langle A^a_\mu(p) A^b_\nu(-p) A^c_\sigma(0) \right\rangle 
&=& -~ ig f^{abc} \left[ ( 2 \eta_{\mu\nu} p_\sigma - 2 \eta_{\mu\sigma} p_\nu 
- 2 \eta_{\nu\sigma} p_\mu ) T_1(p^2) \right. \nonumber \\
&& \left.  ~~~~~~~~~~~~-~
\left[ \eta_{\mu\nu} - \frac{p_\mu p_\nu}{p^2} \right] p_\sigma T_2(p^2)
\right] \nonumber \\
\left\langle c^a(p) \bar{c}^b(-p) A^c_\mu(0) \right\rangle 
&=& -~ ig f^{abc} p_\mu \tilde{\Gamma}_g (p^2) \nonumber \\
\left\langle c^a(0) \bar{c}^b(p) A^c_\mu(-p) \right\rangle 
&=& -~ ig f^{abc} p_\mu \tilde{\Gamma}_c (p^2) \nonumber \\
\left\langle \psi^{iI}(p) \bar{\psi}^{jJ}(0) A^c_\mu(-p) \right\rangle 
&=& -~ g T^c \left[ \gamma_\mu \Lambda_q (p^2) ~+~ \left( \eta_{\mu\nu}
- \frac{p_\mu p_\nu}{p^2} \right) \gamma^\nu \Lambda_q^T(p^2) \right]
\nonumber \\
\left\langle \psi^{iI}(p) \bar{\psi}^{jJ}(-p) A^c_\mu(0) \right\rangle 
&=& -~ g T^c \left[ \gamma_\mu \Lambda_g (p^2) ~+~ \left( \eta_{\mu\nu}
- \frac{p_\mu p_\nu}{p^2} \right) \gamma^\nu \Lambda_g^T(p^2) \right]
\label{greenmomdef}
\end{eqnarray}
where the various form factors are used for the different renormalization
prescriptions and $T^a$ and $f^{abc}$ are the colour group generators and
structure constants respectively. We note that the external momentum 
configuration that defines $\tilde{\Gamma}_c(p^2)$ is the one that is the 
foundation for the $\mMOM$ scheme of \cite{43}.

To define the set of $\MOMtstars$ schemes that we will determine to five loops 
we first recall those of \cite{47} and introduce our syntax. In \cite{47} the
three $\MOMts$ schemes were defined with respect to one nullification of each
of the $3$-point vertices. For each case we record the condition on the 
respective form factors and the $\MOMtstars$ scheme label we will use as its 
notation. We have
\begin{eqnarray}
T_1(\mu^2) ~=~ 1 &\longleftrightarrow& ~ \MOMtgggzgs \nonumber \\
\tilde{\Gamma}_c(\mu^2) ~=~ 1 &\longleftrightarrow& ~ \MOMtccgzcs \nonumber \\
\Lambda_q(\mu^2) ~=~ 1 &\longleftrightarrow& ~ \MOMtqqgzqs 
\label{prescrip1}
\end{eqnarray}
where $g$, $c$ and $q$ denote the gluon, ghost and quark respectively with the
letter after $0$ in the subscript indicating which leg is nullified. In 
\cite{49} the additional scheme introduced in \cite{50} was also examined. In 
the syntax of (\ref{prescrip1}) its defining condition and label is
\begin{equation}
T_1(\mu^2) ~-~ \half T_2(\mu^2) ~=~ 1 ~\longleftrightarrow~ ~ \MOMtgggzggs ~.
\end{equation}
Given that combinations of form factors can be used to form renormalization
prescriptions we introduce the remaining schemes considered here. These are
\begin{eqnarray}
\tilde{\Gamma}_g(\mu^2) &=& 1 ~\longleftrightarrow~ ~ \MOMtccgzgs \nonumber \\
\Lambda_g(\mu^2) &=& 1 ~\longleftrightarrow~ ~ \MOMtqqgzgs \nonumber \\
\Lambda_q(\mu^2) ~+~ \Lambda_q^T(\mu^2) 
&=& 1 ~\longleftrightarrow~ ~ \MOMtqqgzqTs \nonumber \\
\Lambda_g(\mu^2) ~+~ \Lambda_g^T(\mu^2) 
&=& 1 ~\longleftrightarrow~ ~ \MOMtqqgzgTs
\end{eqnarray}
where the equality is the value of the form factors after renormalization. In 
essence these schemes differ from those of \cite{51,52} in that there is only 
one independent momentum flowing through the Green's function instead of two. 
So in effect the computation of the vertex functions reduces to that of
evaluating $2$-point functions. With regard to the prescription for the field
renormalization we first note that the propagators will have the form
\begin{eqnarray}
\left\langle A^a_\mu(p) A^b_\nu(-p) \right\rangle &=& -~
\frac{\delta^{ab}}{p^2} \left[ \left[ \eta_{\mu\nu} - \frac{p_\mu p_\nu}{p^2} 
\right] \frac{1}{[1+\Pi_g(p^2)]} ~-~ \alpha \frac{p_\mu p_\nu}{p^2} \right]
\nonumber \\
\left\langle c^a(p) \bar{c}^b(-p) \right\rangle &=& 
\frac{\delta^{ab}}{p^2} \frac{1}{[1+\Pi_c(p^2)]} \nonumber \\
\left\langle \psi^{iI}(p) \bar{\psi}^{jJ}(-p) \right\rangle &=& 
\frac{\delta^{ij} \delta^{IJ}}{p^2} \frac{\pslash}{[1+\Sigma_q(p^2)]}
\end{eqnarray}
which defines the various $2$-point form factors. The $\MOMtstars$ prescription
is that the propagator denominators are unity at the subtraction point of
$p^2$~$=$~$\mu^2$. As we will also renormalize the quark mass operator, that
immediately determines the quark mass renormalization constant, from the 
Green's function
\begin{equation}
\left\langle \psi^{iI}(p) \bar{\psi}^{jJ}(-p) \,
[ \bar{\psi}^{kK} \psi^{kK} ](0) \right\rangle ~=~ \delta^{ij} \delta^{IJ}
\Gamma_m(p^2) 
\label{massopdef}
\end{equation}
which defines $\Gamma_m(p^2)$ with the prescription that this has to be unity 
at $p^2$~$=$~$\mu^2$. The quark mass renormalization can be treated this way 
using (\ref{massopdef}) as in effect this equates to the way the mass term of 
the quark $2$-point function in the massive version of (\ref{lag}) is 
renormalized. For the massive Lagrangian the quark $2$-point function can be 
expanded in powers of the quark mass $m$ to $O(m)$ thence producing the same
Feynman graphs that constitute (\ref{massopdef}). This is the reason why we 
will only consider the momentum routing in (\ref{massopdef}) rather than the
one where a momentum passes through the operator itself. The key advantage of 
considering (\ref{massopdef}) is that {\sc Forcer} can be exploited to evaluate
the constituent Feynman integrals since the setup will then be a massless one.

One observation needs to be made concerning each scheme. If for instance 
$\left\langle A^a_\mu(p) A^b_\nu(-p) A^c_\sigma(0) \right\rangle$ is selected 
to construct the $\MOMtgggzgs$ scheme renormalization group functions this 
means that $T_1(\mu^2)$ is unity as noted in (\ref{prescrip1}). One question 
then concerns what form do the remaining vertex functions of 
(\ref{greenmomdef}) take at their indicated momentum configuration in the 
$\MOMtgggzgs$ scheme. By the Slavnov-Taylor identities each of the other form
factors will be finite in the same way that using one vertex to find the 
$\MSbar$ coupling renormalization constant means the other vertices are 
immediately finite. However what is generally the case is that each of the 
other vertex functions in the $\MOMtgggzgs$ scheme will not be solely the 
renormalized coupling constant at $p^2$~$=$~$\mu^2$, unless the Slavnov-Taylor 
identities produce this. Instead the form factors of each of the other vertex 
functions will be a perturbative series in the renormalized coupling constant 
of that $\MOMtgggzgs$ scheme. The same observation applies when each of the 
other schemes is considered in turn and this summarizes the comments made in 
\cite{51,52} for the symmetric point MOM schemes. 

Having defined the suite of $\MOMtstars$ schemes in relation to the respective
Green's functions we need to record the particular algorithm to deduce the 
renormalization constants. Assuming the renormalization has been carried out to
$L$ loops in one of the $\MOMtstars$ schemes then the $2$-point functions are 
renormalized at $(L+1)$ loops by removing the poles in $\epsilon$ and the 
finite part. Once the $2$-point functions are rendered finite to $(L+1)$ loops 
then only the form factor of the vertex function of (\ref{greenmomdef}) 
relevant to defining that specific $\MOMtstars$ scheme has its poles and finite
part absorbed into the coupling constant renormalization constant. After this 
has been achieved the process is repeated up to and including four loops 
using the automatic process introduced in \cite{10}. Then this procedure is 
repeated for the subsequent $\MOMtstars$ scheme until all of the schemes have 
been constructed at {\em four} loops. The final step is to determine the 
conversion functions and parameter maps before applying (\ref{confunmap}) to 
deduce the five loop renormalization group functions in each of the 
$\MOMtstars$ schemes. Comment needs to be made on the definition of the 
$\MOMtccgzcs$ scheme in relation to the $\mMOM$ scheme of \cite{43}. That 
scheme is also based on the vertex with an external ghost leg nullification, 
\cite{43}, and leads to the natural question are both schemes equivalent. The 
brief answer is that they are different except in one special instance which is
the Landau gauge. The general difference is because the coupling constant 
renormalization in the $\mMOM$ scheme is constructed from the renormalization 
of $2$-point functions and the $\MSbar$ coupling renormalization constant 
alone, where the finite parts of the $2$-point functions in $\mMOM$ are 
absorbed into the ghost and gluon renormalization constants. Then the $\mMOM$ 
coupling constant renormalization constant is deduced from the relation, 
\cite{43}, 
\begin{equation}
Z_g^{\mMOMss} \sqrt{Z_A^{\mMOMss}} Z_c^{\mMOMss} ~=~
Z_g^{\MSbarss} \sqrt{Z_A^{\MSbarss}} Z_c^{\MSbarss}
\label{mMOMdef}
\end{equation}
which is motivated from Taylor's observation that the ghost-gluon vertex is
finite in the Landau gauge, \cite{41}. For a linear covariant gauge this is
straightforward to see. Irrespective of the renormalization scheme that is
used, for the case where the external ghost nullification is non-trivial the 
vertex containing that ghost has the loop momentum flowing through the other 
ghost and gluon. From (\ref{lag}) the Feynman rule for the vertex involves the 
loop momentum alone with an index contracted with the gluon propagator. This 
immediately reduces the combination to a term proportional to $\alpha$ which 
clearly vanishes in the Landau gauge. The condition (\ref{mMOMdef}), which 
equates the overall ghost-gluon vertex renormalization constants of the two 
schemes, seeks to preserve that equivalence property within the renormalization
for arbitrary $\alpha$. One interesting feature of the four and five loop 
$\mMOM$ $\beta$-function and other renormalization group functions is that they
have neither $\zeta_4$ nor $\zeta_6$ dependence in the Landau 
gauge\footnote{We are indebted to I. Jack for drawing our attention to this 
implicit feature.} which is the special instance noted earlier. There is 
$\zeta_4$ and $\zeta_6$ dependence for non-zero $\alpha$ in the $\mMOM$ scheme.
Indeed this was a clue to the fact that if the constraint for this external 
momentum configuration was ignored and an independent coupling renormalization 
constant was introduced then it should be the case that the $\zeta_4$ and 
$\zeta_6$ dependence is absent in the $\MOMtccgzcs$ scheme for all $\alpha$. In
other words the residual $\zeta_4$ and $\zeta_6$ dependence in terms involving 
$\alpha$, that the condition (\ref{mMOMdef}) omits through its aim of 
preserving the no-renormalization property of the vertex for non-zero $\alpha$,
is accommodated within the $\MOMtccgzcs$ scheme for all $\alpha$. This turned 
out to be indeed the case as will be evident in our results. In one sense the 
Landau gauge $\mMOM$ $\beta$-function could be regarded as another example of 
an AD theory in the classification of \cite{46}.

With the formalism and $\MOMtstars$ scheme definitions established the task of 
determining the renormalization group functions explicitly remains. To do so at
five loops will therefore require a full four loop renormalization of the 
$2$-point functions and the respective Green's functions with the momenta 
configurations of (\ref{greenmomdef}). Helpfully useable data is already 
available, \cite{63}, which is readily extracted using {\sc Forcer}, 
\cite{23,24}. In \cite{63} the $\epsilon$ expansion of the three $2$-point 
functions and the form factors of the three $3$-point functions in 
(\ref{greenmomdef}) have been provided as a function of the bare coupling 
constant and gauge parameter. For the determination of the quark mass 
renormalization constant we use the analogous four loop Green's function that 
was computed in \cite{65}. We note that those four loop expressions had all 
been established with the Feynman graph integration package {\sc Forcer}. To 
extract the respective renormalization constants from each bare Green's 
function we follow the process outlined in \cite{10} where the counterterms are
introduced automatically using the relations given in (\ref{zdef}). Invaluable 
in carrying out the renormalization was the symbolic manipulation language 
{\sc Form}, \cite{25,26}, which meant the extraction of the renormalization 
constants progressed efficiently and these were straightforwardly converted to 
the five loop $\MOMtstars$ renormalization group functions.

\sect{Results.}

One of the main tasks was to demonstrate that in the suite of $\MOMtstars$
schemes no $\zeta_4$ or $\zeta_6$ appears in the renormalization group 
functions. In order to present that observation we have to record the full five
loop expressions that allows one to verify that no such numbers are present. As 
the full expressions for non-zero $\alpha$ and arbitrary colour group are 
rather lengthy we will record the renormalization group functions for one 
scheme as an example and focus on the case of $SU(3)$ and the Landau gauge. 
Except that to illustrate the absence of $\zeta_4$ and $\zeta_6$ for all
$\alpha$ we will record one representative $\beta$-function. Electronic 
versions of the results for all $\alpha$ and a general colour group are 
provided in the data file available via the arXiv version of this paper. It is 
then a simple matter to check that there are neither $\zeta_4$ nor $\zeta_6$ 
terms in any of the renormalization group functions for all the schemes in that
file by employing a search tool for instance. The particular scheme we present 
the results for is the $\MOMtgggzgs$ scheme. First the $\beta$-function for 
non-zero $\alpha$ is
\begin{eqnarray}
\beta^{SU(3)}_{\MOMtgggzgss}(a,\alpha) &=&
\left[
\frac{2}{3} \Nf
- 11
\right] a^2
+ \left[
- \frac{9}{4} \alpha^3
+ \frac{38}{3} \Nf
+ \frac{117}{4} \alpha
- 102
- 3 \Nf \alpha
- \Nf \alpha^2
+ 3 \alpha^2
\right] a^3
\nonumber \\
&&
+ \left[
- \frac{58491}{16}
- \frac{1575}{32} \alpha^3
- \frac{729}{8} \zeta_3 \alpha
- \frac{481}{27} \Nf^2
- \frac{429}{4} \Nf \alpha
- \frac{243}{8} \zeta_3 \alpha^2
\right. \nonumber \\
&& \left. ~~~~
- \frac{165}{8} \Nf \alpha^2
- \frac{119}{6} \zeta_3 \Nf
- \frac{63}{32} \alpha^4
- \frac{8}{9} \zeta_3 \Nf^2
+ \frac{3}{8} \Nf \alpha^4
+ \frac{15}{4} \Nf \alpha^3
+ \frac{1053}{16} \alpha^2
\right. \nonumber \\
&& \left. ~~~~
+ \frac{2277}{4} \zeta_3
+ \frac{5643}{8} \alpha
+ \frac{15283}{24} \Nf
\right] a^4
\nonumber \\
&&
+ \left[
- \frac{10982273}{64}
- \frac{2677587}{128} \zeta_3 \alpha
- \frac{1075423}{144} \zeta_3 \Nf
- \frac{724445}{324} \Nf^2
- \frac{506541}{64} \Nf \alpha
\right. \nonumber \\
&& \left. ~~~~
- \frac{465651}{256} \alpha^3
- \frac{266373}{128} \zeta_3 \alpha^2
- \frac{77441}{64} \Nf \alpha^2
- \frac{60895}{9} \zeta_5 \Nf
- \frac{50481}{256} \alpha^4
\right. \nonumber \\
&& \left. ~~~~
- \frac{8935}{32} \zeta_5 \Nf \alpha
- \frac{5445}{128} \zeta_5 \alpha^4
- \frac{1053}{128} \alpha^5
- \frac{945}{32} \zeta_5 \alpha^5
- \frac{605}{4} \zeta_3 \Nf^2 \alpha
\right. \nonumber \\
&& \left. ~~~~
- \frac{295}{32} \zeta_5 \Nf \alpha^3
- \frac{105}{16} \zeta_5 \Nf \alpha^4
- \frac{16}{3} \zeta_3 \Nf^2 \alpha^2
- \frac{16}{9} \zeta_3 \Nf^3
- \frac{9}{8} \Nf \alpha^5
- \frac{8}{9} \Nf^3 \alpha
\right. \nonumber \\
&& \left. ~~~~
+ \frac{32}{9} \zeta_3 \Nf^3 \alpha
+ \frac{39}{8} \zeta_3 \Nf \alpha^4
+ \frac{75}{8} \Nf \alpha^4
+ \frac{351}{16} \zeta_3 \alpha^5
+ \frac{445}{9} \Nf^2 \alpha^2
\right. \nonumber \\
&& \left. ~~~~
+ \frac{529}{32} \zeta_3 \Nf \alpha^3
+ \frac{788}{27} \Nf^3
+ \frac{3709}{16} \zeta_3 \Nf \alpha^2
+ \frac{9280}{27} \zeta_5 \Nf^2
+ \frac{11903}{48} \Nf^2 \alpha
\right. \nonumber \\
&& \left. ~~~~
+ \frac{12237}{64} \Nf \alpha^3
+ \frac{12959}{54} \zeta_3 \Nf^2
+ \frac{17271}{128} \zeta_3 \alpha^4
+ \frac{20505}{128} \zeta_5 \alpha^3
+ \frac{36135}{2} \zeta_5
\right. \nonumber \\
&& \left. ~~~~
+ \frac{53097}{128} \zeta_3 \alpha^3
+ \frac{82249}{32} \zeta_3 \Nf \alpha
+ \frac{155205}{128} \zeta_5 \alpha^2
+ \frac{251793}{64} \alpha^2
+ \frac{830955}{128} \zeta_5 \alpha
\right. \nonumber \\
&& \left. ~~~~
+ \frac{1425171}{32} \zeta_3
+ \frac{2540673}{64} \alpha
+ \frac{3830167}{96} \Nf
\right] a^5
\nonumber \\
&&
+ \left[
- \frac{65313445615}{27648} \zeta_5 \Nf
- \frac{25790811345}{2048}
- \frac{18051846813}{4096} \zeta_7
\right. \nonumber \\
&& \left. ~~~~
- \frac{6588808297}{10368} \zeta_3 \Nf
- \frac{3473740893}{8192} \zeta_7 \alpha
- \frac{3264898519}{10368} \Nf^2
+ \frac{5165077893}{2048} \alpha
\right. \nonumber \\
&& \left. ~~~~
+ \frac{15665964165}{2048} \zeta_5
+ \frac{22912517957}{18432} \zeta_7 \Nf
+ \frac{910029320399}{248832} \Nf
\right. \nonumber \\
&& \left. ~~~~
- \frac{1356316467}{1024} \zeta_3 \alpha
- \frac{650074537}{1024} \Nf \alpha
- \frac{303580683}{8192} \zeta_7 \alpha^2
\right. \nonumber \\
&& \left. ~~~~
- \frac{242578035}{4096} \zeta_5 \alpha^2
- \frac{224554437}{2048} \alpha^3
- \frac{128139255}{4096} \zeta_5 \alpha^3
- \frac{51105929}{864} \zeta_7 \Nf^2
\right. \nonumber \\
&& \left. ~~~~
- \frac{31554195}{4096} \zeta_5 \alpha^4
- \frac{29533401}{1024} \zeta_3^2 \alpha
- \frac{19040275}{256} \Nf \alpha^2
- \frac{17251731}{4096} \zeta_7 \alpha^3
\right. \nonumber \\
&& \left. ~~~~
- \frac{8416485}{8192} \zeta_7 \alpha^5
- \frac{3996279}{2048} \zeta_7 \alpha^4
- \frac{3120849}{512} \alpha^4
- \frac{2401857}{256} \zeta_3^2 \Nf
\right. \nonumber \\
&& \left. ~~~~
- \frac{1232293}{128} \zeta_3 \Nf \alpha^3
- \frac{1096745}{384} \zeta_3 \Nf \alpha^2
- \frac{721445}{576} \zeta_5 \Nf^2 \alpha^2
- \frac{586025}{108} \zeta_5 \Nf^3
\right. \nonumber \\
&& \left. ~~~~
- \frac{436751}{24} \zeta_3 \Nf^2 \alpha
- \frac{314631}{1024} \alpha^5
- \frac{279765}{1024} \zeta_5 \alpha^5
- \frac{234087}{2048} \zeta_7 \Nf \alpha^4
\right. \nonumber \\
&& \left. ~~~~
- \frac{137619}{1024} \zeta_3 \alpha^6
- \frac{120423}{256} \zeta_3 \Nf \alpha^4
- \frac{103743}{1024} \zeta_3 \alpha^5
- \frac{80865}{1024} \zeta_5 \alpha^6
\right. \nonumber \\
&& \left. ~~~~
- \frac{53411}{36} \zeta_3^2 \Nf^2
- \frac{18823}{36} \Nf^3 \alpha
- \frac{16627}{24} \Nf^2 \alpha^3
- \frac{13455}{8} \zeta_5 \Nf^2 \alpha
\right. \nonumber \\
&& \left. ~~~~
- \frac{7653}{128} \zeta_3 \Nf \alpha^5
- \frac{7065}{64} \Nf \alpha^5
- \frac{4263}{4} \zeta_7 \Nf^2 \alpha
- \frac{2565}{128} \alpha^6
- \frac{2482}{27} \Nf^4
\right. \nonumber \\
&& \left. ~~~~
- \frac{806}{9} \Nf^3 \alpha^2
- \frac{710}{9} \zeta_5 \Nf^3 \alpha
- \frac{409}{8} \Nf^2 \alpha^4
- \frac{304}{27} \zeta_3 \Nf^4
- \frac{245}{4} \zeta_3^2 \Nf^2 \alpha
\right. \nonumber \\
&& \left. ~~~~
- \frac{143}{3} \zeta_3 \Nf^3 \alpha^2
- \frac{117}{32} \zeta_3 \Nf \alpha^6
- \frac{32}{3} \zeta_3 \Nf^3 \alpha^3
- \frac{15}{2} \zeta_5 \Nf^2 \alpha^3
- \frac{9}{16} \Nf \alpha^6
\right. \nonumber \\
&& \left. ~~~~
+ \frac{5}{3} \zeta_5 \Nf^3 \alpha^2
+ \frac{8}{3} \Nf^3 \alpha^3
+ \frac{315}{64} \zeta_5 \Nf \alpha^6
+ \frac{423}{8} \zeta_3^2 \Nf \alpha^3
+ \frac{621}{64} \zeta_3^2 \Nf \alpha^5
\right. \nonumber \\
&& \left. ~~~~
+ \frac{1141}{2} \zeta_3 \Nf^2 \alpha^3
+ \frac{1225}{3} \zeta_3 \Nf^3 \alpha
+ \frac{1760}{27} \zeta_5 \Nf^4
+ \frac{2240}{27} \zeta_3^2 \Nf^3
+ \frac{3549}{1024} \zeta_7 \Nf \alpha^5
\right. \nonumber \\
&& \left. ~~~~
+ \frac{3615}{128} \zeta_5 \Nf \alpha^5
+ \frac{6027}{128} \zeta_3^2 \Nf \alpha^4
+ \frac{27945}{512} \zeta_3^2 \alpha^6
+ \frac{28805}{256} \zeta_5 \Nf \alpha^4
\right. \nonumber \\
&& \left. ~~~~
+ \frac{46845}{256} \zeta_3^2 \Nf \alpha^2
+ \frac{61715}{64} \zeta_5 \Nf \alpha^3
+ \frac{90297}{128} \Nf \alpha^4
+ \frac{159705}{8192} \zeta_7 \alpha^6
\right. \nonumber \\
&& \left. ~~~~
+ \frac{172179}{512} \zeta_3^2 \alpha^5
+ \frac{351521}{288} \zeta_3 \Nf^2 \alpha^2
+ \frac{410595}{128} \zeta_3^2 \Nf \alpha
+ \frac{441851}{648} \zeta_3 \Nf^3
\right. \nonumber \\
&& \left. ~~~~
+ \frac{470223}{1024} \zeta_3^2 \alpha^4
+ \frac{605583}{1024} \zeta_3^2 \alpha^3
+ \frac{1673385}{1024} \zeta_7 \Nf \alpha^2
+ \frac{2574009}{1024} \zeta_3^2 \alpha^2
\right. \nonumber \\
&& \left. ~~~~
+ \frac{5911227}{1024} \zeta_3 \alpha^4
+ \frac{10147923}{512} \Nf \alpha^3
+ \frac{16512415}{512} \zeta_5 \Nf \alpha
+ \frac{25235989}{2592} \Nf^3
\right. \nonumber \\
&& \left. ~~~~
+ \frac{30395757}{1024} \zeta_3 \alpha^2
+ \frac{36466719}{128} \zeta_3 \Nf \alpha
+ \frac{39704059}{6912} \Nf^2 \alpha^2
+ \frac{42976673}{1152} \Nf^2 \alpha
\right. \nonumber \\
&& \left. ~~~~
+ \frac{43537571}{1024} \zeta_7 \Nf \alpha
+ \frac{73833111}{1024} \zeta_3 \alpha^3
+ \frac{74629145}{3072} \zeta_5 \Nf \alpha^2
\right. \nonumber \\
&& \left. ~~~~
+ \frac{87682327}{5184} \zeta_3 \Nf^2
+ \frac{95256513}{512} \zeta_3^2
+ \frac{384102657}{128} \zeta_3
+ \frac{399957783}{2048} \alpha^2
\right. \nonumber \\
&& \left. ~~~~
+ \frac{985401955}{5184} \zeta_5 \Nf^2
+ \frac{1171309995}{4096} \zeta_5 \alpha
- 5 \zeta_3^2 \Nf^2 \alpha^2
+ 15 \zeta_3 \Nf^2 \alpha^4
\right] a^6 \nonumber \\
&& +~ O(a^7) ~.
\label{betasu3ggg0g}
\end{eqnarray}
Our convention throughout will be that the variables $a$ and $\alpha$ are in 
the scheme attached to the renormalization group function itself. The Landau 
gauge anomalous dimensions of the gluon, ghost, quark and quark mass in the 
same scheme are
\begin{eqnarray}
\gamma^{SU(3)}_{A,\MOMtgggzgss}(a,0) &=&
\left[
- \frac{13}{2}
+ \frac{2}{3} \Nf
\right] a
+ \left[
- \frac{255}{4}
+ \frac{67}{6} \Nf
\right] a^2
\nonumber \\
&&
+~ \left[
- \frac{68433}{32}
- \frac{373}{27} \Nf^2
- \frac{83}{6} \zeta_3 \Nf
- \frac{8}{9} \zeta_3 \Nf^2
+ \frac{4365}{16} \zeta_3
+ \frac{11293}{24} \Nf
\right] a^3
\nonumber \\
&&
+~ \left[
- \frac{6789623}{64}
- \frac{1895585}{288} \zeta_5 \Nf
- \frac{1683719}{288} \zeta_3 \Nf
- \frac{1140955}{648} \Nf^2
- \frac{16}{9} \zeta_3 \Nf^3
\right. \nonumber \\
&& \left. ~~~~
+ \frac{671}{27} \Nf^3
+ \frac{9280}{27} \zeta_5 \Nf^2
+ \frac{20707}{108} \zeta_3 \Nf^2
+ \frac{1090305}{64} \zeta_5
+ \frac{2793877}{96} \Nf
\right. \nonumber \\
&& \left. ~~~~
+ \frac{7071297}{256} \zeta_3
\right] a^4
\nonumber \\
&&
+~ \left[
- \frac{74174308299}{16384} \zeta_7
- \frac{67523852455}{27648} \zeta_5 \Nf
- \frac{33044274229}{4096}
\right. \nonumber \\
&& \left. ~~~~
- \frac{7482447221}{20736} \zeta_3 \Nf
+ \frac{2602029675}{2048} \zeta_3
+ \frac{7832693295}{1024} \zeta_5
\right. \nonumber \\
&& \left. ~~~~
+ \frac{24152007425}{18432} \zeta_7 \Nf
+ \frac{658685409635}{248832} \Nf
- \frac{311150303}{1296} \Nf^2
\right. \nonumber \\
&& \left. ~~~~
- \frac{51407573}{864} \zeta_7 \Nf^2
- \frac{3552969}{256} \zeta_3^2 \Nf
- \frac{590285}{108} \zeta_5 \Nf^3
- \frac{54527}{36} \zeta_3^2 \Nf^2
\right. \nonumber \\
&& \left. ~~~~
- \frac{1906}{27} \Nf^4
- \frac{304}{27} \zeta_3 \Nf^4
+ \frac{1760}{27} \zeta_5 \Nf^4
+ \frac{2240}{27} \zeta_3^2 \Nf^3
+ \frac{545495}{648} \zeta_3 \Nf^3
\right. \nonumber \\
&& \left. ~~~~
+ \frac{14365339}{5184} \zeta_3 \Nf^2
+ \frac{19696993}{2592} \Nf^3
+ \frac{660827493}{2048} \zeta_3^2
\right. \nonumber \\
&& \left. ~~~~
+ \frac{1002897415}{5184} \zeta_5 \Nf^2
\right] a^5 ~+~ O(a^6) 
\end{eqnarray}
\begin{eqnarray}
\gamma^{SU(3)}_{c,\MOMtgggzgss}(a,0) &=&
-~ \frac{9}{4} a
+ \left[
- \frac{153}{8}
+ \frac{3}{4} \Nf
\right] a^2
\nonumber \\
&&
+~ \left[
- \frac{46305}{64}
- \frac{5}{2} \Nf^2
+ \frac{9}{4} \zeta_3 \Nf
+ \frac{357}{4} \Nf
+ \frac{1971}{32} \zeta_3
\right] a^3
\nonumber \\
&&
+~ \left[
- \frac{8486505}{256}
- \frac{13205}{48} \Nf^2
- \frac{5895}{16} \zeta_5 \Nf
- \frac{1977}{4} \zeta_3 \Nf
+ \frac{5}{2} \Nf^3
+ \frac{162405}{32} \zeta_5
\right. \nonumber \\
&& \left. ~~~~
+ \frac{748593}{128} \Nf
+ \frac{1824363}{512} \zeta_3
+ 26 \zeta_3 \Nf^2
\right] a^4
\nonumber \\
&&
+~ \left[
- \frac{18930657429}{8192}
- \frac{17060330889}{32768} \zeta_7
+ \frac{5463760635}{4096} \zeta_5
- \frac{435383505}{4096} \zeta_3^2
\right. \nonumber \\
&& \left. ~~~~
- \frac{98927995}{2304} \Nf^2
- \frac{74407755}{512} \zeta_5 \Nf
- \frac{18046313}{512} \zeta_3 \Nf
- \frac{3969}{2} \zeta_7 \Nf^2
\right. \nonumber \\
&& \left. ~~~~
- \frac{219}{2} \zeta_3^2 \Nf^2
+ \frac{7747}{6} \Nf^3
+ \frac{109901}{48} \zeta_3 \Nf^2
+ \frac{282185}{48} \zeta_5 \Nf^2
\right. \nonumber \\
&& \left. ~~~~
+ \frac{846441}{128} \zeta_3^2 \Nf
+ \frac{38085719}{1024} \zeta_7 \Nf
+ \frac{421898185}{768} \Nf
\right. \nonumber \\
&& \left. ~~~~
+ \frac{735148629}{4096} \zeta_3
- 65 \zeta_5 \Nf^3
- 14 \Nf^4
+ \zeta_3 \Nf^3
\right] a^5 ~+~ O(a^6) 
\end{eqnarray}
\begin{eqnarray}
\gamma^{SU(3)}_{\psi,\MOMtgggzgss}(a,0) &=&
\left[
- \frac{4}{3} \Nf
+ \frac{67}{3}
\right] a^2
+ \left[
- \frac{706}{9} \Nf
- \frac{607}{2} \zeta_3
+ \frac{8}{9} \Nf^2
+ \frac{29675}{36}
+ 16 \zeta_3 \Nf
\right] a^3
\nonumber \\
&&
+~ \left[
- \frac{21455183}{648} \zeta_3
- \frac{2401655}{324} \Nf
- \frac{272}{9} \zeta_3 \Nf^2
- \frac{40}{9} \Nf^3
+ \frac{2879}{9} \Nf^2
\right. \nonumber \\
&& \left. ~~~~
+ \frac{73873}{27} \zeta_3 \Nf
+ \frac{7727771}{162}
+ \frac{15846715}{1296} \zeta_5
- 830 \zeta_5 \Nf
\right] a^4
\nonumber \\
&&
+~ \left[
- \frac{93917679073}{31104} \zeta_3
- \frac{26588447977}{27648} \zeta_7
+ \frac{2239174289}{10368} \zeta_3^2
+ \frac{9255603625}{3888} \zeta_5
\right. \nonumber \\
&& \left. ~~~~
+ \frac{14692571119}{5184}
- \frac{333206965}{972} \zeta_5 \Nf
- \frac{252766199}{432} \Nf
- \frac{4731481}{243} \zeta_3 \Nf^2
\right. \nonumber \\
&& \left. ~~~~
- \frac{576437}{54} \zeta_3^2 \Nf
- \frac{80840}{81} \Nf^3
- \frac{6811}{2} \zeta_7 \Nf^2
- \frac{3520}{27} \zeta_5 \Nf^3
- \frac{128}{9} \zeta_3^2 \Nf^2
\right. \nonumber \\
&& \left. ~~~~
+ \frac{160}{27} \Nf^4
+ \frac{24064}{81} \zeta_3 \Nf^3
+ \frac{1069795}{27} \Nf^2
+ \frac{3085750}{243} \zeta_5 \Nf^2
+ \frac{4429579}{36} \zeta_7 \Nf
\right. \nonumber \\
&& \left. ~~~~
+ \frac{1741317151}{3888} \zeta_3 \Nf
\right] a^5 ~+~ O(a^6) 
\end{eqnarray}
and
\begin{eqnarray}
\gamma^{SU(3)}_{m,\MOMtgggzgss}(a,0) &=&
-~ 4 a
+ \left[
- \frac{209}{3}
+ \frac{4}{3} \Nf
\right] a^2
\nonumber \\
&&
+~ \left[
- \frac{23731}{9}
- \frac{176}{9} \zeta_3 \Nf
- \frac{8}{3} \Nf^2
+ \frac{2723}{3} \zeta_3
+ \frac{4823}{27} \Nf
\right] a^3
\nonumber \\
&&
+~ \left[
- \frac{364717295}{2592}
- \frac{318905}{54} \zeta_3 \Nf
- \frac{16015}{3} \zeta_5
- \frac{13741}{27} \Nf^2
- \frac{3200}{9} \zeta_5 \Nf
\right. \nonumber \\
&& \left. ~~~~
+ \frac{8}{3} \Nf^3
+ \frac{1552}{9} \zeta_3 \Nf^2
+ \frac{5306821}{324} \Nf
+ \frac{30602221}{432} \zeta_3
\right] a^4
\nonumber \\
&&
+~ \left[
- \frac{1206873493973}{124416}
- \frac{2433059707}{3456} \zeta_3^2
+ \frac{22567052305}{3888} \zeta_3
\right. \nonumber \\
&& \left. ~~~~
+ \frac{77544762803}{46656} \Nf
+ \frac{105673656865}{124416} \zeta_5
- \frac{2023786561}{2592} \zeta_3 \Nf
\right. \nonumber \\
&& \left. ~~~~
- \frac{870630095}{7776} \zeta_5 \Nf
- \frac{362933429}{3888} \Nf^2
- \frac{68062141}{1296} \zeta_7 \Nf
- \frac{60928}{81} \zeta_3^2 \Nf^2
\right. \nonumber \\
&& \left. ~~~~
- \frac{28096}{81} \zeta_3 \Nf^3
- \frac{1600}{9} \zeta_5 \Nf^3
- \frac{352}{27} \Nf^4
+ \frac{1372}{3} \zeta_7 \Nf^2
+ \frac{468142}{243} \Nf^3
\right. \nonumber \\
&& \left. ~~~~
+ \frac{3766661}{108} \zeta_3 \Nf^2
+ \frac{3825215}{486} \zeta_5 \Nf^2
+ \frac{6552685}{162} \zeta_3^2 \Nf
+ \frac{657118063}{27648} \zeta_7
\right] a^5 \nonumber \\
&& +~ O(a^6) 
\end{eqnarray}
respectively. As an alternative perspective where the $\zeta_n$ structure is
equally evident it is instructive to view the Landau gauge pure Yang-Mills
theory results for an arbitrary colour group. We have
\begin{eqnarray}
\left. \beta_{\MOMtgggzgss}(a,0) \right|_{\Nf = 0} &=&
-~ \frac{11}{3} C_A a^2
- \frac{34}{3} C_A^2 a^3
+ \left[
- \frac{6499}{48} C_A^3
+ \frac{253}{12} \zeta_3 C_A^3
\right] a^4
\nonumber \\
&&
+~ \left[
- \frac{10981313}{5184} C_A^4
- \frac{3707}{8} \zeta_3 \frac{d_A^{abcd} d_A^{abcd}}{\NA} 
- \frac{8}{9} \frac{d_A^{abcd} d_A^{abcd}}{\NA}
\right. \nonumber \\
&& \left. ~~~~
+ \frac{6215}{24} \zeta_5 \frac{d_A^{abcd} d_A^{abcd}}{\NA}
+ \frac{97405}{576} \zeta_5 C_A^4
+ \frac{1116929}{1728} \zeta_3 C_A^4
\right] a^5
\nonumber \\
&&
+~ \left[
- \frac{8598255605}{165888} C_A^5
- \frac{1161130663}{73728} \zeta_7 C_A^5
- \frac{35208635}{3072} \zeta_7 C_A \frac{d_A^{abcd} d_A^{abcd}}{\NA}
\right. \nonumber \\
&& \left. ~~~~
- \frac{28905223}{2304} \zeta_3 C_A \frac{d_A^{abcd} d_A^{abcd}}{\NA}
- \frac{15922907}{9216} \zeta_3^2 C_A^5
\right. \nonumber \\
&& \left. ~~~~
+ \frac{131849}{3456} C_A \frac{d_A^{abcd} d_A^{abcd}}{\NA}
+ \frac{4595789}{384} \zeta_3^2 C_A \frac{d_A^{abcd} d_A^{abcd}}{\NA}
\right. \nonumber \\
&& \left. ~~~~
+ \frac{7284505}{1152} \zeta_5 C_A \frac{d_A^{abcd} d_A^{abcd}}{\NA}
+ \frac{30643529}{2048} \zeta_3 C_A^5
\right. \nonumber \\
&& \left. ~~~~
+ \frac{1667817635}{55296} \zeta_5 C_A^5
\right] a^6 ~+~ O(a^7)
\nonumber \\
\left. \gamma_{A,\MOMtgggzgss}(a,0) \right|_{\Nf = 0} &=&
-~ \frac{13}{6} C_A a
- \frac{85}{12} C_A^2 a^2
+ \left[
- \frac{22811}{288} C_A^3
+ \frac{485}{48} \zeta_3 C_A^3
\right] a^3
\nonumber \\
&&
+~ \left[
- \frac{13595371}{10368} C_A^4
- \frac{45245}{192} \zeta_5 \frac{d_A^{abcd} d_A^{abcd}}{\NA}
- \frac{475}{32} \zeta_3 \frac{d_A^{abcd} d_A^{abcd}}{\NA}
\right. \nonumber \\
&& \left. ~~~~
+ \frac{1075}{144} \frac{d_A^{abcd} d_A^{abcd}}{\NA}
+ \frac{1189237}{3456} \zeta_3 C_A^4
+ \frac{1195385}{4608} \zeta_5 C_A^4
\right] a^4
\nonumber \\
&&
+~ \left[
- \frac{33074782019}{995328} C_A^5
- \frac{15073026227}{884736} \zeta_7 C_A^5
+ \frac{2673449615}{82944} \zeta_5 C_A^5
\right. \nonumber \\
&& \left. ~~~~
- \frac{282030679}{36864} \zeta_7 C_A \frac{d_A^{abcd} d_A^{abcd}}{\NA}
- \frac{129543941}{110592} \zeta_3^2 C_A^5
\right. \nonumber \\
&& \left. ~~~~
- \frac{6255185}{1728} \zeta_5 C_A \frac{d_A^{abcd} d_A^{abcd}}{\NA}
- \frac{4876861}{768} \zeta_3 C_A \frac{d_A^{abcd} d_A^{abcd}}{\NA}
\right. \nonumber \\
&& \left. ~~~~
+ \frac{3050779}{20736} C_A \frac{d_A^{abcd} d_A^{abcd}}{\NA}
+ \frac{55278899}{4608} \zeta_3^2 C_A \frac{d_A^{abcd} d_A^{abcd}}{\NA}
\right. \nonumber \\
&& \left. ~~~~
+ \frac{543400985}{82944} \zeta_3 C_A^5
\right] a^5 ~+~ O(a^6) 
\nonumber \\
\left. \gamma_{c,\MOMtgggzgss}(a,0) \right|_{\Nf = 0} &=&
-~ \frac{3}{4} C_A a
- \frac{17}{8} C_A^2 a^2
+ \left[
- \frac{1715}{64} C_A^3
+ \frac{73}{32} \zeta_3 C_A^3
\right] a^3
\nonumber \\
&&
+~ \left[
- \frac{26235}{64} C_A^4
- \frac{3375}{64} \zeta_3 \frac{d_A^{abcd} d_A^{abcd}}{\NA}
+ \frac{101}{32} \frac{d_A^{abcd} d_A^{abcd}}{\NA}
\right. \nonumber \\
&& \left. ~~~~
+ \frac{6795}{128} \zeta_5 \frac{d_A^{abcd} d_A^{abcd}}{\NA}
+ \frac{7037}{128} \zeta_3 C_A^4
+ \frac{52835}{1024} \zeta_5 C_A^4
\right] a^4
\nonumber \\
&&
+~ \left[
- \frac{2110649851}{221184} C_A^5
- \frac{161581091}{65536} \zeta_7 C_A^5
- \frac{4327735}{24576} \zeta_3^2 C_A^5
\right. \nonumber \\
&& \left. ~~~~
- \frac{1979089}{384} \zeta_3 C_A \frac{d_A^{abcd} d_A^{abcd}}{\NA}
- \frac{1284495}{1024} \zeta_3^2 C_A \frac{d_A^{abcd} d_A^{abcd}}{\NA}
\right. \nonumber \\
&& \left. ~~~~
+ \frac{80483}{512} C_A \frac{d_A^{abcd} d_A^{abcd}}{\NA}
+ \frac{4450325}{768} \zeta_5 C_A \frac{d_A^{abcd} d_A^{abcd}}{\NA}
\right. \nonumber \\
&& \left. ~~~~
+ \frac{12700107}{8192} \zeta_7 C_A \frac{d_A^{abcd} d_A^{abcd}}{\NA}
+ \frac{66809507}{36864} \zeta_3 C_A^5
\right. \nonumber \\
&& \left. ~~~~
+ \frac{157858255}{36864} \zeta_5 C_A^5
\right] a^5 ~+~ O(a^6) 
\end{eqnarray}
where higher order colour Casimirs arise with the fully symmetric rank $4$ 
tensor defined by, \cite{66},
\begin{equation}
d_R^{abcd} ~=~ \frac{1}{6} \mbox{Tr} \left( T^a T^{(b} T^c T^{d)}
\right)
\end{equation}
in representation $R$. In expressions with these Casimirs either here or in the
associated data file we have implemented the identity $\Nc C_F$~$=$~$T_F \NA$. 
We note that the pure Yang-Mills expressions for the other schemes have a 
similar structure. 

There are several checks on the results. The main one is that we have verified
that the Landau gauge four loop $\beta$-functions of the $\MOMtccgzcs$,
$\MOMtqqgzqs$, $\MOMtgggzgs$ and $\MOMtgggzggs$ schemes agree with the $SU(3)$
expressions given in \cite{49}. In addition we have reproduced the $\alpha$ 
dependent $O(a^4)$ coupling constant mappings for the same schemes as those 
recorded in \cite{49} for the same colour group at four loops. There is another
check on our computations which is the five loop QED $\beta$-function provided 
in \cite{53} for the $\MOMts$ scheme although it was termed the MOM scheme 
there. In \cite{53} QED was renormalized at five loops in the $\MSbar$ scheme. 
The $\MOMts$ $\beta$-function was produced as a corollary via the 
Ward-Takahashi identity. This implies that the coupling constant and photon 
renormalization constants are not independent placing the theory in the AD 
class of \cite{46}. So the $\MOMts$ scheme $\beta$-function of \cite{53} 
follows immediately by ensuring the photon $2$-point function has its finite 
part absorbed into its renormalization constant. Taking the QED limit of the 
$\MOMtqqgzgTs$ scheme renormalization group functions reproduces the 
$\beta$-function of \cite{53}. We note this gives  
\begin{eqnarray}
\beta_{\MOMteepzpTss}^{\QEDss}(a,\alpha) &=&
\frac{4 \Nf}{3} a^2 + 4 \Nf a^3
+ [ 96 \zeta_3 \Nf - 92 \Nf - 9 ] \frac{2 \Nf a^4}{9}
\nonumber \\
&&
+~ [ - 128 \zeta_3 \Nf^2 + 192 \Nf^2 + 256 \zeta_3 \Nf - 640 \zeta_5 \Nf 
+ 156 \Nf - 69 ] \frac{2 \Nf a^5}{3}
\nonumber \\
&&
+~ [ 21504 \zeta_3 \Nf^3 + 30720 \zeta_5 \Nf^3 - 51456 \Nf^3 
+ 55296 \zeta_3^2 \Nf^2 - 157440 \zeta_3 \Nf^2 
\nonumber \\
&& ~~~~
+ 138240 \zeta_5 \Nf^2 
- 54128 \Nf^2 - 52992 \zeta_3 \Nf - 311040 \zeta_5 \Nf + 483840 \zeta_7 \Nf 
\nonumber \\
&& ~~~~
- 54216 \Nf + 6912 \zeta_3 + 37413 ] \frac{\Nf a^6}{54} ~+~ O(a^7)
\nonumber \\
\gamma_{A,\MOMteepzpTss}^{\QEDss}(a,\alpha) 
&=& \frac{4 \Nf}{3} a + 4 \Nf a^2  
+ [ 96 \zeta_3 \Nf - 92 \Nf - 9 ] \frac{2 \Nf a^3}{9}
\nonumber \\
&&
+~ [ - 128 \zeta_3 \Nf^2 + 192 \Nf^2 + 256 \zeta_3 \Nf - 640 \zeta_5 \Nf 
+ 156 \Nf - 69 ] \frac{2 \Nf a^4}{3}
\nonumber \\
&&
+~ [ 21504 \zeta_3 \Nf^3 + 30720 \zeta_5 \Nf^3 - 51456 \Nf^3 
+ 55296 \zeta_3^2 \Nf^2 - 157440 \zeta_3 \Nf^2 
\nonumber \\
&& ~~~~
+ 138240 \zeta_5 \Nf^2 
- 54128 \Nf^2 - 52992 \zeta_3 \Nf - 311040 \zeta_5 \Nf + 483840 \zeta_7 \Nf 
\nonumber \\
&& ~~~~
- 54216 \Nf + 6912 \zeta_3 + 37413 ] \frac{\Nf a^5}{54} ~+~ O(a^6)
\nonumber \\
\gamma_{\psi,\MOMteepzpTss}^{\QEDss}(a,\alpha) 
&=& \alpha a - [ 4 \Nf + 3 ] \frac{a^2}{2}
+ [ 16 \Nf^2 - 12 \Nf + 9 ] \frac{a^3}{6} 
\nonumber \\
&&
+~ [ - 128 \alpha \Nf^2 - 96 \alpha \Nf - 640 \Nf^3 - 768 \zeta_3 \Nf^2 
+ 1200 \Nf^2 - 384 \zeta_3 \Nf 
\nonumber \\
&& ~~~~
+ 3304 \Nf - 9600 \zeta_3 + 15360 \zeta_5 
- 3081 ] \frac{a^4}{24} 
\nonumber \\
&&
+~ [ - 768 \zeta_3 \alpha^2 \Nf^2 + 384 \alpha^2 \Nf^2 
- 576 \zeta_3 \alpha^2 \Nf + 288 \alpha^2 \Nf - 3072 \zeta_3 \alpha \Nf^3 
\nonumber \\
&& ~~~~
+ 2304 \alpha \Nf^3 - 2304 \zeta_3 \alpha \Nf^2 - 576 \alpha \Nf^2 
- 432 \alpha \Nf + 5120 \Nf^4 
\nonumber \\
&& ~~~~
+ 28672 \zeta_3 \Nf^3 - 55424 \Nf^3 + 147456 \zeta_3^2 \Nf^2 
- 398336 \zeta_3 \Nf^2 + 496640 \zeta_5 \Nf^2 
\nonumber \\
&& ~~~~
- 162576 \Nf^2 
+ 317952 \zeta_3^2 \Nf - 184128 \zeta_3 \Nf 
+ 1674240 \zeta_5 \Nf 
\nonumber \\
&& ~~~~
- 1979712 \zeta_7 \Nf - 20568 \Nf + 179712 \zeta_3^2 
+ 1152000 \zeta_3 
+ 1627200 \zeta_5 
\nonumber \\
&& ~~~~
- 3429216 \zeta_7 
+ 44793 ] \frac{a^5}{72} ~+~ O(a^6)
\label{qedeep0pT}
\end{eqnarray}
for an arbitrary gauge parameter where we use $e$ and $p$ in the scheme label
to denote the electron and photon respectively in order to be clear which
external leg was nullified. The $\beta$-function of (\ref{qedeep0pT}) does
indeed agree with the $\MOMts$ $\beta$-function of \cite{53} and none of the 
expressions involve $\zeta_4$ or $\zeta_6$. Therefore we confirm that the 
vertex subtraction of \cite{53} corresponds to nullifying the photon of the QED
vertex. We have included the electron anomalous dimension (\ref{qedeep0pT}) as 
it was not present in \cite{53}. Unlike the QCD case the QED $\MOMteepzpTs$ 
$\beta$-function is $\alpha$ independent. We have also reproduced 
(\ref{qedeep0pT}) directly in order to find the parameter mappings which are
\begin{eqnarray}
a^{\QEDss}_{\MOMteepzpTss} &=&
a_{\MSbarss} - \frac{20 \Nf}{9} a_{\MSbarss}^2 
+ [ 400 \Nf + 1296 \zeta_3 - 1485 ] \frac{\Nf}{81} a_{\MSbarss}^3 
\nonumber \\
&&
+~ 2 [ - 4000 \Nf^2 - 50544 \zeta_3 \Nf + 63009 \Nf + 35964 \zeta_3 
- 58320 \zeta_5
+ 11583 ] \frac{\Nf}{729} a_{\MSbarss}^4 
\nonumber \\
&&
+~ [ 320000 \Nf^3 + 11244096 \zeta_3 \Nf^2 + 1866240 \zeta_5 \Nf^2 
- 15967908 \Nf^2 
+ 8957952 \zeta_3^2 \Nf 
\nonumber \\
&& ~~~~
- 33195744 \zeta_3 \Nf 
+ 454896 \zeta_4 \Nf + 28460160 \zeta_5 \Nf 
- 186381 \Nf 
- 1364688 \zeta_3 
\nonumber \\
&& ~~~~
- 25719120 \zeta_5 + 29393280 \zeta_7
+ 135594 ] \frac{\Nf}{13122} a_{\MSbarss}^5 ~+~ O(a_{\MSbarss}^6)
\nonumber \\
\alpha^{\QEDss}_{\MOMteepzpTss} &=&
\alpha_{\MSbarss} + \frac{20\Nf}{9} \alpha_{\MSbarss} a_{\MSbarss} 
+ \alpha_{\MSbarss} [ - 48 \zeta_3 + 55 ] \frac{\Nf}{3} a_{\MSbarss}^2 
\nonumber \\
&&
+~ 2 \alpha_{\MSbarss} [ 2736 \zeta_3 \Nf - 3701 \Nf 
- 3996 \zeta_3 + 6480 \zeta_5 - 1287 ] \frac{\Nf}{81} a_{\MSbarss}^3 
\nonumber \\
&&
+~ \alpha_{\MSbarss} [ - 465984 \zeta_3 \Nf^2 - 207360 \zeta_5 \Nf^2 
+ 786052 \Nf^2 - 622080 \zeta_3^2 \Nf 
\nonumber \\
&& ~~~~~~~~~
+ 2193696 \zeta_3 \Nf 
- 50544 \zeta_4 \Nf - 2125440 \zeta_5 \Nf + 304839 \Nf 
+ 151632 \zeta_3 
\nonumber \\
&& ~~~~~~~~~
+ 2857680 \zeta_5 - 3265920 \zeta_7 
- 15066 ] \frac{\Nf}{1458} a_{\MSbarss}^4 ~+~ O(a_{\MSbarss}^5) ~.
\end{eqnarray}

While we have concentrated on the structure of the renormalization group 
functions of the $\MOMtstars$ schemes the conversion functions for the gluon,
ghost, quark and quark mass share an interesting property which is
\begin{eqnarray}
C^{\MOMtccgzcss}_\phi(a,\alpha) &=& C^{\MOMtccgzgss}_\phi(a,\alpha) ~=~ 
C^{\MOMtgggzgss}_\phi(a,\alpha) \nonumber \\
&=& C^{\MOMtqqgzqss}_\phi(a,\alpha) ~=~ C^{\MOMtqqgzgss}_\phi(a,\alpha) 
\label{confunequiv}
\end{eqnarray}
for each $\phi$ in the same labelling set as previously. These are all derived 
from the renormalization of $2$-point functions with the same subtraction 
condition. This equivalence property equally occurs in the regularization 
invariant ($\RI$) and $\mMOM$ schemes, \cite{64,66}, as well as those 
associated with the symmetric point MOM schemes of \cite{51,52} that were 
provided in \cite{67,68}. The common underlying property of all the 
$C^{\MOMtstarss}_\phi(a,\alpha)$ conversion functions and the corresponding 
$\RI$, $\mMOM$ and MOM ones is that the prescription to define the respective 
wave function and quark mass renormalization constants in each of the schemes 
is the same. In other words the finite part of each $2$-point function is 
absorbed into the renormalization constant. Moreover while the expressions for 
say $Z_A$ constructed with this prescription in two different schemes will be 
formally different, in the determination of their conversion functions with 
respect to the reference $\MSbar$ scheme the effect of the different coupling 
constant and gauge parameter mappings wash out. What is not the case is that 
there is a parallel equivalence for $C_g^{\cal S}(a,\alpha)$ as is evident from
the data associated with the arXiv version of this article. Moreover they ought
not to be since the prescription to define $Z_g$ for each $\MOMtstars$ is 
different. While (\ref{confunequiv}) provides an interesting property of the 
conversion functions it could in principle ease future compilations of 
renormalization group functions for the wave function and quark mass anomalous 
dimensions. In other words for schemes where such $2$-point subtractions are to
be implemented one in effect only requires the coupling constant map to be 
computed explicitly. That for the gauge parameter is not independent of 
$C_A^{\cal S}(a,\alpha)$ in a linear covariant gauge. Furthermore one could 
have schemes which are hybrid in the sense that some $2$-point functions are 
renormalized with an $\MSbar$ prescription whereas the remaining ones are 
rendered finite with a finite subtraction too. In this sense the $\RI$ scheme
of \cite{69,70} could be regarded as a hybrid scheme since the coupling 
constant is renormalized with an $\MSbar$ prescription, meaning 
$C_g^{\RIs}(a,\alpha)$ is trivially and obviously unity, but the $2$-point 
functions have their finite parts subtracted, \cite{47,49,65} and satisfy 
(\ref{confunequiv}). Finally as a side comment the fact that 
(\ref{confunequiv}) was observed at five loops by direct explicit computation 
provides in part a reassuring consistency check on our overall approach.

\sect{$C$-scheme mapping.}

Having established that the $\MOMtstars$ scheme renormalization group functions
do not have any $\zeta_4$ or $\zeta_6$ dependence one question that arises is
whether one of these schemes is in fact equivalent to the $C$-scheme of
\cite{59,60}. One claim of \cite{59,60} is that $\zeta_4$ is absent for certain
physical quantities. One way to test whether there is a connection with the 
$C$-scheme is to compare the $\MOMtstars$ coupling constant maps with the map 
given in equation (7) of \cite{59}. While that depends on the parameter $C$ in 
the order $L$ polynomial of the $O(a^{L+1})$ term of the mapping it might be 
possible to find a particular value of $C$ that exactly matches the mapping of 
an $\MOMtstars$ scheme. Therefore in order to facilitate a comparison with 
\cite{59,60} we note that for $\Nf$~$=$~$3$ the $SU(3)$ mappings in the Landau 
gauge are
\begin{eqnarray}
\left. a_{\MOMtccgzcss} \right|_{\alpha=0,\Nf=3}^{SU(3)} &=& a 
+ \frac{43}{4} a^2
+ \left[
\frac{15685}{48}
- \frac{383}{8} \zeta_3
\right] a^3
\nonumber \\
&&
+ \left[
\frac{20589011}{1728}
- \frac{408251}{144} \zeta_3
- \frac{62255}{192} \zeta_5
\right] a^4
\nonumber \\
&&
+ \left[
\frac{2446354687}{36864} \zeta_7
- \frac{683706835}{4608} \zeta_3
- \frac{606373645}{4608} \zeta_5
+ \frac{3911}{32} \zeta_3^2
\right. \nonumber \\
&& \left. ~~~
+ \frac{627809683}{1152}
+ 1335 \zeta_4
\right] a^5 ~+~ O(a^6) \nonumber \\
\left. a_{\MOMtccgzgss} \right|_{\alpha=0,\Nf=3}^{SU(3)} &=& a
+ \frac{61}{4} a^2
+ \left[
\frac{22597}{48}
- \frac{383}{8} \zeta_3
\right] a^3
\nonumber \\
&&
+ \left[
\frac{15762289}{864}
- \frac{970483}{288} \zeta_3
- \frac{74405}{192} \zeta_5
\right] a^4
\nonumber \\
&&
+ \left[
\frac{15799466317}{18432}
+ \frac{1876891639}{36864} \zeta_7
- \frac{2039209463}{9216} \zeta_3
- \frac{1934713015}{18432} \zeta_5
\right. \nonumber \\
&& \left. ~~~
+ \frac{766871}{512} \zeta_3^2
+ 1335 \zeta_4
\right] a^5 ~+~ O(a^6) \nonumber \\
\left. a_{\MOMtgggzgss} \right|_{\alpha=0,\Nf=3}^{SU(3)} &=& a 
+ \frac{43}{4} a^2
+ \left[
\frac{7973}{24}
- \frac{223}{4} \zeta_3
\right] a^3
+ \left[
\frac{41856073}{3456}
- \frac{1925333}{576} \zeta_3
- \frac{575}{12} \zeta_5
\right] a^4
\nonumber \\
&&
+ \left[
\frac{10225015489}{18432}
- \frac{1369710205}{18432} \zeta_5
- \frac{110437973}{576} \zeta_3
- \frac{131329}{512} \zeta_3^2
\right. \nonumber \\
&& \left. ~~~
+ \frac{1652451493}{36864} \zeta_7
+ 1335 \zeta_4
\right] a^5 ~+~ O(a^6) \nonumber \\
\left. a_{\MOMtgggzggss} \right|_{\alpha=0,\Nf=3}^{SU(3)} &=& a 
+ 16 a^2
+ \left[
\frac{93427}{192}
- \frac{169}{4} \zeta_3
\right] a^3
\nonumber \\
&&
+ \left[
\frac{129114635}{6912}
- \frac{1822913}{576} \zeta_3
- \frac{124835}{192} \zeta_5
\right] a^4
\nonumber \\
&&
+ \left[
\frac{4050665663}{4608}
- \frac{393488663}{2304} \zeta_3
+ \frac{980775}{512} \zeta_3^2
+ \frac{1055749471}{36864} \zeta_7
\right. \nonumber \\
&& \left. ~~~
- \frac{1387483355}{9216} \zeta_5
+ 1335 \zeta_4
\right] a^5 ~+~ O(a^6) \nonumber \\
\left. a_{\MOMtqqgzgss} \right|_{\alpha=0,\Nf=3}^{SU(3)} &=& a
+ \frac{25}{4} a^2
+ \left[
\frac{725}{4}
- 85 \zeta_3
\right] a^3
+ \left[
\frac{127615}{64} \zeta_5
- \frac{542609}{144} \zeta_3
+ \frac{12018703}{3456}
\right] a^4
\nonumber \\
&&
+ \left[
\frac{5829675395}{82944}
+ \frac{2035638385}{41472} \zeta_5
- \frac{523779403}{18432} \zeta_7
- \frac{457075871}{5184} \zeta_3
\right. \nonumber \\
&& \left. ~~~
+ \frac{895703}{128} \zeta_3^2
+ 1335 \zeta_4
\right] a^5 ~+~ O(a^6) \nonumber \\
\left. a_{\MOMtqqgzgTss} \right|_{\alpha=0,\Nf=3}^{SU(3)} &=& a 
+ \frac{37}{4} a^2
+ \left[
\frac{1843}{6}
- 98 \zeta_3
\right] a^3
+ \left[
\frac{36955015}{3456}
- \frac{705631}{144} \zeta_3
+ \frac{199895}{576} \zeta_5
\right] a^4
\nonumber \\
&&
+ \left[
\frac{13618908001}{27648}
+ \frac{2145762283}{55296} \zeta_7
- \frac{1176358325}{13824} \zeta_5
- \frac{296853959}{1152} \zeta_3
\right. \nonumber \\
&& \left. ~~~
+ \frac{28000843}{1152} \zeta_3^2
+ 1335 \zeta_4
\right] a^5 ~+~ O(a^6) \nonumber \\
\left. a_{\MOMtqqgzqss} \right|_{\alpha=0,\Nf=3}^{SU(3)} &=& a 
+ \frac{43}{4} a^2
+ \left[
\frac{16009}{48}
- 58 \zeta_3
\right] a^3
+ \left[
\frac{21116969}{1728}
- \frac{241291}{72} \zeta_3
- \frac{51815}{192} \zeta_5
\right] a^4
\nonumber \\
&&
+ \left[
\frac{1298610053}{2304}
- \frac{829799785}{4608} \zeta_3
- \frac{560109325}{4608} \zeta_5
+ \frac{1068169}{256} \zeta_3^2
\right. \nonumber \\
&& \left. ~~~
+ \frac{1047188135}{18432} \zeta_7
+ 1335 \zeta_4
\right] a^5 ~+~ O(a^6) \nonumber \\
\left. a_{\MOMtqqgzqTss} \right|_{\alpha=0,\Nf=3}^{SU(3)} &=& a
+ \frac{227}{12} a^2
+ \left[
\frac{42365}{72}
- \frac{653}{8} \zeta_3
\right] a^3
\nonumber \\
&&
+ \left[
- \frac{7398635}{1296} \zeta_3
- \frac{3660955}{5184} \zeta_5
+ \frac{62323649}{2592}
\right] a^4
\nonumber \\
&&
+ \left[
\frac{73210930375}{62208}
- \frac{51236375219}{124416} \zeta_3
- \frac{28655962325}{124416} \zeta_5
+ \frac{12942575401}{110592} \zeta_7
\right. \nonumber \\
&& \left. ~~~
+ \frac{922617353}{20736} \zeta_3^2
+ 1335 \zeta_4
\right] a^5 ~+~ O(a^6) 
\label{ccmaps}
\end{eqnarray}
where $a$ on the right hand side is in the $\MSbar$ scheme. In order to 
quantify the behaviour of the mappings the numerical values of (\ref{ccmaps}) 
are
\begin{eqnarray}
\left. a_{\MOMtccgzcss} \right|_{\alpha=0,\Nf=3}^{SU(3)} &=&
a + 10.7500 a^2 + 269.2224 a^3 + 8170.7954 a^4 + 298706.1459 a^5 + O(a^6) \nonumber \\
\left. a_{\MOMtccgzgss} \right|_{\alpha=0,\Nf=3}^{SU(3)} &=&
a + 15.2500 a^2 + 413.22236 a^3 + 13790.9432 a^4 + 537305.8623 a^5 + O(a^6) \nonumber \\
\left. a_{\MOMtgggzgss} \right|_{\alpha=0,\Nf=3}^{SU(3)} &=&
a + 10.7500 a^2 + 265.1937 a^3 + 8043.4603 a^4 + 293487.5638 a^5 + O(a^6) \nonumber \\
\left. a_{\MOMtgggzggss} \right|_{\alpha=0,\Nf=3}^{SU(3)} &=&
a + 16.0000 a^2 + 435.8121 a^3 + 14201.3422 a^4 + 550737.2450 a^5 + O(a^6) \nonumber \\
\left. a_{\MOMtqqgzgss} \right|_{\alpha=0,\Nf=3}^{SU(3)} &=&
a + 6.2500 a^2 + 79.0752 a^3 + 1015.7594 a^4 - 1902.2308 a^5 + O(a^6) \nonumber \\
\left. a_{\MOMtqqgzgTss} \right|_{\alpha=0,\Nf=3}^{SU(3)} &=&
a + 9.2500 a^2 + 189.3651 a^3 + 5162.5198 a^4 + 170286.4367 a^5 + O(a^6) \nonumber \\
\left. a_{\MOMtqqgzqss} \right|_{\alpha=0,\Nf=3}^{SU(3)} &=&
a + 10.7500 a^2 + 263.8015 a^3 + 7912.2228 a^4 + 285890.5240 a^5 + O(a^6) \nonumber \\
\left. a_{\MOMtqqgzqTss} \right|_{\alpha=0,\Nf=3}^{SU(3)} &=&
a + 18.9167 a^2 + 490.2849 a^3 + 16450.0060 a^4 + 626761.5038 a^5 + O(a^6) ~.
\nonumber \\ 
\end{eqnarray}
All bar the mapping for $\MOMtqqgzgs$ have a similar form in the sense that all
the corrections are positive. The four loop coefficient of the $\MOMtqqgzgs$
scheme mapping is negative. Although the gauge used in \cite{59,60} is not 
specified we have chosen the Landau gauge to also indicate some general 
properties of the mappings first. 

For completeness we provide an example of the gauge parameter mapping. Again
choosing $\Nf$~$=$~$3$ for $\alpha$~$\neq$~$0$ we have 
\begin{eqnarray}
\left. \alpha_{\MOMtgggzgss} \right|_{\Nf = 3}^{SU(3)} &=&
\alpha
+ \left[
- \frac{19}{4} \alpha
- \frac{3}{2} \alpha^2
- \frac{3}{4} \alpha^3
\right] a
\nonumber \\
&&
+~ \left[
- \frac{11537}{96} \alpha
- \frac{9}{16} \alpha^4
+ \frac{9}{16} \alpha^5
+ \frac{15}{16} \alpha^3
+ \frac{303}{32} \alpha^2
- 18 \zeta_3 \alpha^2
+ 31 \zeta_3 \alpha
\right] a^2 
\nonumber \\
&&
+~ \left[
- \frac{12861817}{3456} \alpha
- \frac{26517}{32} \zeta_3 \alpha^2
- \frac{8919}{128} \alpha^4
- \frac{6011}{128} \alpha^3
- \frac{1713}{32} \zeta_3 \alpha^3
\right. \nonumber \\
&& \left. ~~~~
- \frac{1341}{32} \zeta_4 \alpha
- \frac{315}{64} \zeta_5 \alpha^5
- \frac{171}{16} \alpha^5
- \frac{45}{8} \zeta_5 \alpha^4
- \frac{27}{64} \alpha^7
+ \frac{27}{16} \alpha^6
+ \frac{81}{8} \zeta_4 \alpha^2
\right. \nonumber \\
&& \left. ~~~~
+ \frac{81}{32} \zeta_4 \alpha^3
+ \frac{117}{32} \zeta_3 \alpha^5
+ \frac{1341}{32} \zeta_3 \alpha^4
+ \frac{3105}{8} \zeta_5 \alpha^2
+ \frac{3465}{32} \zeta_5 \alpha^3
+ \frac{16567}{128} \alpha^2
\right. \nonumber \\
&& \left. ~~~~
+ \frac{41195}{192} \zeta_5 \alpha
+ \frac{82711}{72} \zeta_3 \alpha
\right] a^3 
\nonumber \\
&&
+~ \left[
- \frac{1163178749}{6912} \alpha
- \frac{851518199}{18432} \zeta_7 \alpha
- \frac{87691285}{3072} \zeta_3 \alpha^2
- \frac{60649407}{4096} \zeta_7 \alpha^2
\right. \nonumber \\
&& \left. ~~~~
- \frac{12605197}{4608} \alpha^3
- \frac{4220519}{1536} \zeta_3 \alpha^3
- \frac{2566431}{2048} \zeta_7 \alpha^3
- \frac{1210101}{512} \alpha^4
\right. \nonumber \\
&& \left. ~~~~
- \frac{1075923}{2048} \alpha^5
- \frac{935271}{1024} \zeta_5 \alpha^4
- \frac{851175}{1024} \zeta_6 \alpha^2
- \frac{808011}{2048} \zeta_4 \alpha
\right. \nonumber \\
&& \left. ~~~~
- \frac{209169}{512} \zeta_3^2 \alpha^3
- \frac{203949}{512} \zeta_3^2 \alpha^2
- \frac{137781}{1024} \zeta_7 \alpha^5
- \frac{103875}{32} \zeta_6 \alpha
\right. \nonumber \\
&& \left. ~~~~
- \frac{53865}{256} \zeta_5 \alpha^5
- \frac{52569}{512} \zeta_3^2 \alpha^4
- \frac{23247}{1024} \zeta_4 \alpha^4
- \frac{19755}{512} \zeta_5 \alpha^6
- \frac{17631}{512} \zeta_3 \alpha^6
\right. \nonumber \\
&& \left. ~~~~
- \frac{13041}{2048} \zeta_4 \alpha^5
- \frac{567}{256} \alpha^8
- \frac{351}{64} \zeta_3 \alpha^7
+ \frac{81}{256} \alpha^9
+ \frac{945}{128} \zeta_5 \alpha^7
+ \frac{1863}{256} \zeta_3^2 \alpha^6
\right. \nonumber \\
&& \left. ~~~~
+ \frac{2349}{128} \alpha^7
+ \frac{4725}{1024} \zeta_6 \alpha^5
+ \frac{5373}{512} \alpha^6
+ \frac{10647}{4096} \zeta_7 \alpha^6
+ \frac{19737}{512} \zeta_3^2 \alpha^5
\right. \nonumber \\
&& \left. ~~~~
+ \frac{22275}{1024} \zeta_6 \alpha^3
+ \frac{24975}{1024} \zeta_6 \alpha^4
+ \frac{78327}{512} \zeta_4 \alpha^3
+ \frac{132615}{512} \zeta_7 \alpha^4
+ \frac{205719}{32} \zeta_3^2 \alpha
\right. \nonumber \\
&& \left. ~~~~
+ \frac{558927}{2048} \zeta_3 \alpha^5
+ \frac{663687}{1024} \zeta_4 \alpha^2
+ \frac{1674621}{1024} \zeta_3 \alpha^4
+ \frac{5561003}{1024} \zeta_5 \alpha^3
\right. \nonumber \\
&& \left. ~~~~
+ \frac{8440805}{18432} \alpha^2
+ \frac{25777469}{1024} \zeta_5 \alpha^2
+ \frac{663132857}{9216} \zeta_5 \alpha
+ \frac{891656237}{18432} \zeta_3 \alpha
\right] a^4 \nonumber \\
&& +~ O(a^5) 
\end{eqnarray}
for the $\MOMtgggzgs$ scheme. Unlike the coupling constant map $\zeta_4$ first
appears at $O(a^3)$ and $\zeta_6$ is present at $O(a^4)$. The gauge parameter
maps for the other $\MOMtstars$ schemes are formally the same for all colour 
groups from (\ref{confunequiv}).

One main observation from (\ref{ccmaps}) is that $\zeta_4$ appears in each of 
the $O(a^5)$ terms but $\zeta_6$ is absent. The latter does not arise when 
$\alpha$~$\neq$~$0$ nor for any colour group. In the $O(a^3)$ terms in
(\ref{ccmaps}) $\zeta_3$ is present but in the mapping of \cite{59} there is no
$\zeta_3$ at the same order. Instead there are only rationals. While $\zeta_3$ 
could in principle be introduced by a choice of $C$ that would then mean 
$\zeta_3$ is present in the $O(a^2)$ term which none of the mappings in 
(\ref{ccmaps}) have. Equally if $C$ is determined from the $O(a^2)$ term to 
match that of one of the $O(a^2)$ terms of the $\MOMtstars$ mappings then that 
choice could not introduce a $\zeta_3$ term at $O(a^3)$. At $O(a^4)$ $\zeta_5$ 
is present in (\ref{ccmaps}) but is absent at the corresponding order in the 
map of \cite{59}. By contrast for the $\Nf$~$=$~$3$ expression provided in 
\cite{59} there is a $\zeta_4$ term at $O(a^5)$. Moreover its coefficient is 
{\em precisely} the same as that of $\zeta_4$ at the same order in each of the 
$\MOMtstars$ schemes when $\Nf$~$=$~$3$, after allowing for a factor of $4$ for
differing coupling constant conventions as is evident in (\ref{ccmaps}). It
transpires that this equality occurs for all $\Nf$ and a general colour group.
Moreover it suggests that the coupling constant map of \cite{59,60} does have 
the same underlying $\zeta_4$ cancellation property whatever the 
renormalization prescription that underlies it is. In some sense the 
universality of this particular $\zeta_4$ term in all the mappings reinforces 
the observations of \cite{44,45,46} that the $\zeta_4$ absence could be traced 
to a unique $\epsilon$ dependent transformation of $\zeta_3$. That $\epsilon$ 
dependence would affect the counterterms in the underlying renormalization 
group functions. The fact that there is no other universal connection in any of
the mappings for odd zetas merely reflects the different prescriptions defining
those schemes. We have also examined the situation for $\alpha$~$\neq$~$0$. 
While the extra parameter could in principle be exploited to find a suitable 
value for $C$ to achieve a match at low order, this does not persist at higher 
order. So it would appear that none of our schemes have an immediate connection
to the $C$-scheme aside from the $\zeta_4$ one at $O(a^5)$. 

We can examine the situation from another point of view. It is worth recalling 
the origin of the coupling constant map at a more formal level to see if it 
sheds light on the relation of the $\MOMtstars$ schemes to the $C$-scheme. In 
general the coupling renormalization constants for two schemes take the 
following forms
\begin{equation}
Z_g ~=~ 1 ~+~ \sum_{n=1}^\infty \sum_{m=1}^n z_{g\,nm}
\frac{a^n}{\epsilon^m} ~~~,~~~ 
Z_g^{\cal S} ~=~ 1 ~+~ \sum_{n=1}^\infty \sum_{m=0}^n z_{g\,nm}^{\cal S} 
\frac{a_{\cal S}^n}{\epsilon^m} 
\label{zgdefform}
\end{equation}
where $z_{g\,nm}$ and $z_{g\,nm}^{\cal S}$ are the residues of the poles in 
$\epsilon$ in the $\MSbar$ scheme and a general scheme ${\cal S}$ respectively.
Included in the ${\cal S}$ scheme definition are the finite parts with 
$z_{g\,n0}^{\cal S}$~$\neq$~$0$ but there no corresponding $z_{g\,n0}$ terms in
keeping with the definition of the $\MSbar$ scheme. We will assume there are no
other parameters, such as a gauge parameter, in this formal analysis. So our 
focus is on the Landau gauge. If we define the relation between the two 
coupling constants in perturbation theory as 
\begin{equation}
a_{\cal S} ~=~ \sum_{n=0}^\infty c_n a^{n+1}
\label{amapgen}
\end{equation}
in the same notation as \cite{59} and recall the definition of (\ref{aalmap}) 
which determines the mapping from the coupling renormalization constant, then 
it is a straightforward exercise to deduce
\begin{eqnarray}
c_0 &=& 1 ~~~,~~~
c_1 ~=~ -~ 2 z_{g\,10}^{\cal S} ~~~,~~~
c_2 ~=~ 7 (z_{g\,10}^{\cal S})^2 ~-~ 2 z_{g\,20}^{\cal S} \nonumber \\
c_3 &=& -~ 30 (z_{g\,10}^{\cal S})^3 ~+~ 
18 z_{g\,10}^{\cal S} z_{g\,20}^{\cal S} ~-~ 
2 z_{g\,30}^{\cal S} \nonumber \\
c_4 &=& 143 (z_{g\,10}^{\cal S})^4 ~-~ 
132 (z_{g\,10}^{\cal S})^2 z_{g\,20}^{\cal S} ~+~ 
22 z_{g\,10}^{\cal S} z_{g\,30}^{\cal S} ~+~ 
11 (z_{g\,20}^{\cal S})^2 ~-~ 
2 z_{g\,40}^{\cal S} \nonumber \\
c_5 &=& -~ 728 (z_{g\,10}^{\cal S})^5 ~+~ 
910 (z_{g\,10}^{\cal S})^3 z_{g\,20}^{\cal S} ~-~ 
182 (z_{g\,10}^{\cal S})^2 z_{g\,30}^{\cal S} ~-~ 
182 z_{g\,10}^{\cal S} (z_{g\,20}^{\cal S})^2 \nonumber \\
&& +~ 26 z_{g\,10}^{\cal S} z_{g\,40}^{\cal S} ~+~ 
26 z_{g\,20}^{\cal S} z_{g\,30}^{\cal S} ~-~ 
2 z_{g\,50}^{\cal S} ~. 
\label{ccmapcoeff}
\end{eqnarray}
We note that $z_{g\,nm}$ and $z_{g\,nm}^{\cal S}$ are predetermined when 
$m$~$\geq$~$2$ from the simple pole residues and finite parts of the lower loop
terms at each order $n$ for each renormalization constant. Having formally 
derived (\ref{ccmapcoeff}) we have checked that all the Landau gauge mappings 
of (\ref{ccmaps}) are reproduced from the finite parts computed to four loops. 
In examining the structure of the respective finite parts in each $\MOMtstars$ 
scheme we note that for example $z_{g\,20}^{\cal S}$ involves $\zeta_3$ but 
$z_{g\,10}^{\cal S}$ has only rationals for the $\MOMtstars$ schemes. From 
(\ref{ccmapcoeff}) it is clear that for each $n$ $z_{g\,n0}^{\cal S}$ appears 
for the first time in $c_n$ in addition to all the lower order finite parts. 

If we assume for the moment that the $C$-scheme satisfies these ${\cal S}$
scheme properties we can examine it in more detail. The $C$-scheme involves the
parameter $C$ which was motivated by the observation that the ratio of 
$\Lambda$-parameters between a scheme and the $\MSbar$ scheme is determined 
exactly by $c_1/\beta_1$ where $c_1$ is the one loop term of (\ref{amapgen}) 
and $\beta_1$ is the one loop coefficient of the $O(a^2)$ term in 
(\ref{betasu3ggg0g}). Therefore making this connection with the formal origin 
of $C$ in \cite{59,60}, where $\beta_1$~$=$~$-$~$9$ for 
$\Nf$~$=$~$\Nc$~$=$~$3$, the power series dependence of $z_{g\,10}^{\cal S}$ in
$c_n$ in (\ref{ccmapcoeff}) parallels that of the parameter $C$ in the coupling
constant mapping of \cite{59,60}. Specifically one can check that $C$ is $c_1$ 
which is related to $z_{g\,10}^{\cal S}$. However, the very assumption in 
\cite{59,60} that $c_1$ is non-zero in the $\Lambda$ ratio immediately implies 
that whatever the renormalization prescription is that defines the $C$-scheme, 
at the level of subtracting divergences of a vertex function, it is one where 
$Z_g^{\cal S}$ has a non-zero finite contribution at each loop order. Therefore
there ought to be the equivalent of $z_{g\,n0}^{\cal S}$ dependence in the 
coupling constant map of \cite{59,60} from the $C$-scheme to the $\MSbar$ 
scheme for $n$~$\geq$~$2$. Such dependence appears to be absent as the mapping 
of \cite{59,60} depends on only one parameter and therefore only 
$z_{g\,10}^{\cal S}$. Unless the explicit values of $z_{g\,n0}^{\cal S}$ are 
all zero when computed for all $n$~$\geq$~$2$, which would be peculiar, then it
would appear that it is not possible to connect the $C$-scheme to any of the 
$\MOMtstars$ schemes using the renormalization group based argument that led to
(\ref{ccmapcoeff}). Indeed we took the values of $c_n$ given in \cite{59,60} 
and solved for $z_{g\,n0}^{\cal S}$ for each $\MOMtstars$ scheme. After 
matching $C$ to $c_1$ for each scheme the remaining $z_{g\,n0}^{\cal S}$ for 
$n$~$\geq$~$2$ are not in agreement with the finite parts determined from each 
$\MOMtstars$ renormalization.

Returning to the more general scheme ${\cal S}$ when there is a finite part in 
the coupling renormalization constant of (\ref{zgdefform}), it is instructive 
to record the form of the $\beta$-function for non-zero $\epsilon$ and 
therefore clarify earlier comments. Using (\ref{rgedef}) and (\ref{zgdefform}) 
we have the formal $\epsilon$ dependent $\beta$-function
\begin{eqnarray}
\beta_{\cal S}(a,\epsilon) &=&
2 z^{\cal S}_{g\,11} a^2 
+ 4 \left[ 
- 3 z^{\cal S}_{g\,11} z^{\cal S}_{g\,10} 
+ z^{\cal S}_{g\,21}
\right] a^3 
\nonumber \\
&&
+~ 2 \left[
27 z^{\cal S}_{g\,11} (z^{\cal S}_{g\,10})^2 
- 11 z^{\cal S}_{g\,11} z^{\cal S}_{g\,20} 
- 11 z^{\cal S}_{g\,10} z^{\cal S}_{g\,21} 
+ 3 z^{\cal S}_{g\,31}
\right] a^4 
\nonumber \\
&&
+~ 8 \left[ 
- 27 z^{\cal S}_{g\,11} (z^{\cal S}_{g\,10})^3 
+ 24 z^{\cal S}_{g\,11} z^{\cal S}_{g\,10} z^{\cal S}_{g\,20} 
- 4 z^{\cal S}_{g\,11} z^{\cal S}_{g\,30} 
+ 12 (z^{\cal S}_{g\,10})^2 z^{\cal S}_{g\,21} 
\right. \nonumber \\
&& \left. ~~~~~~
- 4 z^{\cal S}_{g\,10} z^{\cal S}_{g\,31} 
- 5 z^{\cal S}_{g\,20} z^{\cal S}_{g\,21} 
+ z^{\cal S}_{g\,41}
\right] a^5 
\nonumber \\
&&
+~ 2 \left[
405 z^{\cal S}_{g\,11} (z^{\cal S}_{g\,10})^4 
- 567 z^{\cal S}_{g\,11} (z^{\cal S}_{g\,10})^2 z^{\cal S}_{g\,20} 
+ 138 z^{\cal S}_{g\,11} z^{\cal S}_{g\,10} z^{\cal S}_{g\,30} 
+ 85 z^{\cal S}_{g\,11} (z^{\cal S}_{g\,20})^2 
\right. \nonumber \\
&& \left. ~~~~~~
- 21 z^{\cal S}_{g\,11} z^{\cal S}_{g\,40} 
- 189 (z^{\cal S}_{g\,10})^3 z^{\cal S}_{g\,21} 
+ 69 (z^{\cal S}_{g\,10})^2 z^{\cal S}_{g\,31} 
+ 170 z^{\cal S}_{g\,10} z^{\cal S}_{g\,20} z^{\cal S}_{g\,21} 
\right. \nonumber \\
&& \left. ~~~~~~
- 21 z^{\cal S}_{g\,10} z^{\cal S}_{g\,41} 
- 29 z^{\cal S}_{g\,20} z^{\cal S}_{g\,31} 
- 29 z^{\cal S}_{g\,21} z^{\cal S}_{g\,30} 
+ 5 z^{\cal S}_{g\,51}
\right] a^6 
\nonumber \\
&&
+ \left[
-~ a 
+ 2 z^{\cal S}_{g\,10} a^2 
+ 2 \left[ - 3 (z^{\cal S}_{g\,10})^2 
+ 2 z^{\cal S}_{g\,20}
\right] a^3 
\right. \nonumber \\
&& \left. ~~~
+~ 2 \left[
9 (z^{\cal S}_{g\,10})^3 
- 11 z^{\cal S}_{g\,10} z^{\cal S}_{g\,20} 
+ 3 z^{\cal S}_{g\,30}
\right] a^4 
\right. \nonumber \\
&& \left. ~~~
+~ 2 \left[ 
- 27 (z^{\cal S}_{g\,10})^4 
+ 48 (z^{\cal S}_{g\,10})^2 z^{\cal S}_{g\,20} 
- 16 z^{\cal S}_{g\,10} z^{\cal S}_{g\,30} 
- 10 (z^{\cal S}_{g\,20})^2 
+ 4 z^{\cal S}_{g\,40}
\right] a^5 
\right. \nonumber \\
&& \left. ~~~
+~ 2 \left[
81 (z^{\cal S}_{g\,10})^5 
- 189 (z^{\cal S}_{g\,10})^3 z^{\cal S}_{g\,20} 
+ 69 (z^{\cal S}_{g\,10})^2 z^{\cal S}_{g\,30} 
+ 85 z^{\cal S}_{g\,10} (z^{\cal S}_{g\,20})^2 
\right. \right. \nonumber \\
&& \left. \left. ~~~~~~~~~~
- 21 z^{\cal S}_{g\,10} z^{\cal S}_{g\,40} 
- 29 z^{\cal S}_{g\,20} z^{\cal S}_{g\,30} 
+ 5 z^{\cal S}_{g\,50}
\right] a^6 \right] \epsilon ~+~ O(a^7) 
\label{betaS}
\end{eqnarray}
where the $\epsilon$ dependent contributions follow the part that survives when
the regularization is lifted. A similar expression can be constructed for the
anomalous dimension of the fields and mass. In each case the coefficients will
depend not only on the residues and finite parts of the respective 
renormalization constants but also on $Z_g^{\cal S}$. We recall that in a gauge
theory the corresponding construction will be more involved for a non-zero 
covariant gauge parameter. Like (\ref{ccmapcoeff}) the $O(\epsilon)$ term of
$\beta_{\cal S}(a)$ depends solely on $z^{\cal S}_{g\,n0}$ for $n$~$\geq$~$1$.
So knowledge of the coefficients of $a$ in either of these means the 
coefficients of $a$ in the other can be determined. 

The necessity of the $O(\epsilon)$ piece is central to another aspect of the
renormalization group properties. This concerns critical exponents which are
renormalization group invariants and given by the evaluation of the anomalous
dimensions at zeros of the $\beta$-function. In the case of the latter the
relevant exponent is the slope of the $\beta$-function at criticality. Amongst
the widely studied suite of exponents are those derived from the Wilson-Fisher
fixed point, \cite{71}, defined as the critical point closest to the origin for
non-zero $\epsilon$. For the $\MSbar$ scheme, where there are no $\epsilon$
terms in the $\beta$-function aside from the $O(a)$ one, which itself reflects 
the dimensionlessness of the $d$-dimensional coupling constant, the exponent
$\omega$~$=$~$\beta_{\cal S}^\prime(a,\epsilon)$ where the derivative acts on
$a$, will only depend on the residues of the simple poles of $\epsilon$ in 
$Z_g^{\MSbarss}$ as is clear from (\ref{betaS}). Equally (\ref{betaS}) suggests
that evaluating $\omega$ for the generic scheme ${\cal S}$ would involve 
$z^{\cal S}_{g\,n0}$ as well. This might seem to imply that $\omega$ would be 
different in different schemes and hence contradict the renormalization group 
invariance of the exponents at the Wilson-Fisher fixed point. We have checked 
this is not the case to $O(\epsilon^5)$ in each of the $\MOMtstars$ schemes 
considered here. This was for the Landau gauge as that is a fixed point of 
$\gamma_\alpha(a,\alpha)$. Moreover, the agreement has also been verified in 
\cite{72} for the MOM schemes of \cite{51,52}. In other words for a generic 
scheme the invariance of the exponents actually provides relations between 
$z_{g\,n1}$ and $z^{\cal S}_{g\,n1}$. In particular we record
\begin{eqnarray}
z^{\cal S}_{g\,21} &=&
3 z_{g\,11} z^{\cal S}_{g\,10}
+ z_{g\,21} ~~~,~~~
z^{\cal S}_{g\,31} ~=~
3 z_{g\,11} (z^{\cal S}_{g\,10})^2
+ 3 z_{g\,11} z^{\cal S}_{g\,20}
+ 5 z_{g\,21} z^{\cal S}_{g\,10}
+ z_{g\,31} 
\nonumber \\
z^{\cal S}_{g\,41} &=&
z_{g\,11} (z^{\cal S}_{g\,10})^3
+ 6 z_{g\,11} z^{\cal S}_{g\,10} z^{\cal S}_{g\,20}
+ 3 z_{g\,11} z^{\cal S}_{g\,30}
+ 10 z_{g\,21} (z^{\cal S}_{g\,10})^2
+ 5 z_{g\,21} z^{\cal S}_{g\,20}
+ 7 z_{g\,31} z^{\cal S}_{g\,10}
+ z_{g\,41} 
\nonumber \\
z^{\cal S}_{g\,51} &=&
3 z_{g\,11} (z^{\cal S}_{g\,10})^2 z^{\cal S}_{g\,20}
+ 6 z_{g\,11} z^{\cal S}_{g\,10} z^{\cal S}_{g\,30}
+ 3 z_{g\,11} (z^{\cal S}_{g\,20})^2
+ 3 z_{g\,11} z^{\cal S}_{g\,40}
+ 10 z_{g\,21} (z^{\cal S}_{g\,10})^3
\nonumber \\
&&
+~ 20 z_{g\,21} z^{\cal S}_{g\,10} z^{\cal S}_{g\,20}
+ 5 z_{g\,21} z^{\cal S}_{g\,30}
+ 21 z_{g\,31} (z^{\cal S}_{g\,10})^2
+ 7 z_{g\,31} z^{\cal S}_{g\,20}
+ 9 z_{g\,41} z^{\cal S}_{g\,10}
+ z_{g\,51} 
\end{eqnarray}
where we have assumed $z^{\cal S}_{g\,11}$~$=$~$z_{g\,11}$. While the two loop
term of the $\beta$-function in a single coupling theory is scheme independent
this does not imply $z_{g\,21}$ and $z^{\cal S}_{g\,21}$ are equal. Although we
have checked these relations are satisfied in the $\MOMtstars$ schemes we 
cannot do the same for the $C$-scheme $\beta$-function as only the purely four 
dimensional expression is available and not the $\epsilon$ dependent one. For 
the wave function renormalization $Z_\phi$ similar relations hold between the
term of the respective renormalization constants. If we define 
\begin{equation}
Z_\phi ~=~ 1 ~+~ \sum_{n=1}^\infty \sum_{m=1}^n z_{\phi\,nm}
\frac{a^n}{\epsilon^m} ~~~,~~~ 
Z_\phi^{\cal S} ~=~ 1 ~+~ \sum_{n=1}^\infty \sum_{m=0}^n z_{\phi\,nm}^{\cal S} 
\frac{a_{\cal S}^n}{\epsilon^m} 
\label{zpdefform}
\end{equation}
it is straightforward to deduce
\begin{eqnarray}
z^{\cal S}_{\phi\,21} &=&
2 z^{\cal S}_{g\,10} z_{\phi\,11}
+ z_{\phi\,11} z^{\cal S}_{\phi\,10}
+ z_{\phi\,21} 
\nonumber \\
z^{\cal S}_{\phi\,31} &=&
(z^{\cal S}_{g\,10})^2 z_{\phi\,11}
+ 2 z^{\cal S}_{g\,10} z_{\phi\,11} z^{\cal S}_{\phi\,10}
+ 4 z^{\cal S}_{g\,10} z_{\phi\,21}
+ 2 z^{\cal S}_{g\,20} z_{\phi\,11}
+ z_{\phi\,11} z^{\cal S}_{\phi\,20}
+ z_{\phi\,21} z^{\cal S}_{\phi\,10}
+ z_{\phi\,31} 
\nonumber \\
z^{\cal S}_{\phi\,41} &=&
(z^{\cal S}_{g\,10})^2 z_{\phi\,11} z^{\cal S}_{\phi\,10}
+ 6 (z^{\cal S}_{g\,10})^2 z_{\phi\,21}
+ 2 z^{\cal S}_{g\,10} z^{\cal S}_{g\,20} z_{\phi\,11}
+ 2 z^{\cal S}_{g\,10} z_{\phi\,11} z^{\cal S}_{\phi\,20}
+ 4 z^{\cal S}_{g\,10} z_{\phi\,21} z^{\cal S}_{\phi\,10}
\nonumber \\
&&
+~ 6 z^{\cal S}_{g\,10} z_{\phi\,31}
+ 2 z^{\cal S}_{g\,20} z_{\phi\,11} z^{\cal S}_{\phi\,10}
+ 4 z^{\cal S}_{g\,20} z_{\phi\,21}
+ 2 z^{\cal S}_{g\,30} z_{\phi\,11}
+ z_{\phi\,11} z^{\cal S}_{\phi\,30}
+ z_{\phi\,21} z^{\cal S}_{\phi\,20}
\nonumber \\
&&
+~ z_{\phi\,31} z^{\cal S}_{\phi\,10}
+ z_{\phi\,41} 
\nonumber \\
z^{\cal S}_{\phi\,51} &=&
4 (z^{\cal S}_{g\,10})^3 z_{\phi\,21}
+ (z^{\cal S}_{g\,10})^2 z_{\phi\,11} z^{\cal S}_{\phi\,20}
+ 6 (z^{\cal S}_{g\,10})^2 z_{\phi\,21} z^{\cal S}_{\phi\,10}
+ 15 (z^{\cal S}_{g\,10})^2 z_{\phi\,31}
\nonumber \\
&&
+~ 2 z^{\cal S}_{g\,10} z^{\cal S}_{g\,20} z_{\phi\,11} z^{\cal S}_{\phi\,10}
+ 12 z^{\cal S}_{g\,10} z^{\cal S}_{g\,20} z_{\phi\,21}
+ 2 z^{\cal S}_{g\,10} z^{\cal S}_{g\,30} z_{\phi\,11}
+ 2 z^{\cal S}_{g\,10} z_{\phi\,11} z^{\cal S}_{\phi\,30}
\nonumber \\
&&
+~ 4 z^{\cal S}_{g\,10} z_{\phi\,21} z^{\cal S}_{\phi\,20}
+ 6 z^{\cal S}_{g\,10} z_{\phi\,31} z^{\cal S}_{\phi\,10}
+ 8 z^{\cal S}_{g\,10} z_{\phi\,41}
+ (z^{\cal S}_{g\,20})^2 z_{\phi\,11}
+ 2 z^{\cal S}_{g\,20} z_{\phi\,11} z^{\cal S}_{\phi\,20}
\nonumber \\
&&
+~ 4 z^{\cal S}_{g\,20} z_{\phi\,21} z^{\cal S}_{\phi\,10}
+ 6 z^{\cal S}_{g\,20} z_{\phi\,31}
+ 2 z^{\cal S}_{g\,30} z_{\phi\,11} z^{\cal S}_{\phi\,10}
+ 4 z^{\cal S}_{g\,30} z_{\phi\,21}
+ 2 z^{\cal S}_{g\,40} z_{\phi\,11}
+ z_{\phi\,11} z^{\cal S}_{\phi\,40}
\nonumber \\
&&
+~ z_{\phi\,21} z^{\cal S}_{\phi\,30}
+ z_{\phi\,31} z^{\cal S}_{\phi\,20}
+ z_{\phi\,41} z^{\cal S}_{\phi\,10}
+ z_{\phi\,51} 
\end{eqnarray}
where $z^{\cal S}_{\phi\,11}$~$=$~$z_{\phi\,11}$ has been assumed.

\sect{Perspective on schemes.}

Having completed the explict construction of the $\MOMtstars$ schemes at
five loops in QCD it is worth pausing to consider the position of such schemes
in a more general context. The discussion, however, will be for massless 
theories so that particle masses do not feature in the underlying
renormalization. First for the moment we will focus on a theory with a single 
field and an $n$-point interaction. Although initially we will consider a 
$3$-point interaction as it will provide a simple introduction to classes of 
schemes. For instance, we will suggest that for such an interaction there are 
two classes of schemes which will be termed $1$- and $3$-variable. By 
$1$-variable we mean those schemes where there is only one independent 
invariant or equivalently variable, which for the $\MOMts$ prescription is the 
external momentum of the $2$- and $3$-point functions that are used to 
determine the renormalization constants and recorded as an example in 
(\ref{greenmomdef}) for QCD. For the vertex function the momentum configuration
is an exceptional one in that there are fewer independent momenta than the 
maximum permitted for a $3$-point function; for an $n$-point function this is 
$(n-1)$. While the structure of the Feynman rules for each of the three 
vertices in QCD ensures that infrared rearrangement trivially implies there are
no infrared issues it also means that the number basis of the nullified 
$3$-point vertices is $\zeta_n$ or multiple zetas up to at least six loops 
aside from rationals. For $\phi^3$ theory in six dimensions one can nullify the
external momentum of the cubic vertex since that is automatically infrared safe
unlike four dimensions. By contrast with this $1$-variable notion of scheme the 
$3$-variable scheme for a cubic theory corresponds to schemes where the vertex 
momentum configuration is non-exceptional. In this instance the number basis is
known to be different from that of the $1$-variable case. For such $3$-point 
function configurations there will be more invariants and hence they will 
depend on several variables. The reason for this is that there are now two 
independent momenta. For instance, for a $3$-point vertex with non-zero 
external momenta $p_1$, $p_2$ and $p_3$ one of these is not independent, say 
$p_3$, via energy-momentum conservation. From the two independent momenta there
are three scalar products $p_1^2$, $p_2^2$ and $p_1.p_2$ which can be regarded 
as two scales and essentially an angle. Alternatively one could take $p_i^2$ 
for $i$~$=$~$1$, $2$ and $3$ as the independent set of variables. For the 
$3$-point function one can form two dimensionless variables 
$x$~$=$~$\frac{p_1^2}{p_3^2}$ and $y$~$=$~$\frac{p_2^2}{p_3^2}$, say, leaving 
one variable $p_3^2$ as the dimensionful one. The overall scale will be common 
to all the Feynman graphs comprising a $3$-point vertex meaning that the
remaining vertex function, prior to renormalization, will depend on $x$ and $y$
which are not renormalized. By contrast, in a $1$-variable scheme the variable 
itself, which is the square of the external momentum, does not feature in the 
actual renormalization constants purely as it is dimensionful but it will be
present in the finite renormalized Green's function. So there is no remnant of 
the kinematics of the subtraction configuration in the renormalization group
functions in a $1$-variable scheme unlike the $3$-variable one in this cubic 
theory example.

An example of such a scenario is the well-established symmetric point momentum 
subtraction scheme defined in QCD in \cite{51,52}. For instance the full 
renormalization group functions of QCD in the three MOM schemes are available
at three loops in \cite{51,52,67,68,73} in an arbitrary linear covariant gauge 
and at four loops in the Landau gauge in \cite{68}. In other words these three 
MOM schemes have $x$~$=$~$y$~$=$~$1$. However there is no a priori reason why 
$x$ and $y$ should take these values. In principle they can be left as free 
variables although restricted to configurations where there are no collinear or
infrared singularities for example. While schemes with $x$ and $y$ both free 
have not been studied as such at the Lagrangian level, a subset of non-unit $x$
and $y$ values have been, not only for the Lagrangian but also for operator 
renormalization in what is termed the interpolating MOM scheme, 
\cite{74,75,76}. This is the case where $x$ and $y$ are related to one common 
parameter $\hat{\omega}$\footnote{In \cite{74,75,76} the variable is actually 
$\omega$ but we use $\hat{\omega}$ briefly here to avoid confusion with the use
of $\omega$ for a different entity later in this section.}. Considering a 
$3$-variable scheme which depends on two variables may seem an irrelevant 
exercise but it could have the advantage of tuning the convergence for the 
perturbative series of an observable to minimize theory uncertainties or 
alternatively provide a more informed method of estimating theory errors for 
instance. Equally having schemes depend on variables may not be aesthetically 
pleasing. On the other hand there is no a priori reason why the symmetric point 
configuration $x$~$=$~$y$~$=$~$1$ should be singled out for special 
significance. A related issue to tuning is the situation of taking mathematical
limits. One such limit would be that which should produce the lower variable 
scheme or schemes which for a cubic theory would be the $1$-variable one. 
Therefore one could regard the development of the $\MOMtstars$ schemes in QCD 
as both the starting point as well as an endpoint check for building such a 
suite of schemes.

Of course in the QCD example the Lagrangian also possesses a quartic gluon
vertex which in this vision would lead to another class of schemes. The pattern 
for this is now clear and would be a set of $6$-variable schemes with five 
dimensionless variables. More generally for an $n$-point function the
non-exceptional scheme would be an $\half n(n-1)$-variable one. In the case of 
$n$~$=$~$4$ with momenta $p_1$, $p_2$, $p_3$ and $p_4$ we can take the first
three as the independent ones by energy-momentum conservation which can be used
to construct six invariants. For example, these could either be the lengths of 
the six possible momenta, $p_1$, $p_2$, $p_3$, $(p_1+p_2)$, $(p_1+p_3)$ and 
$(p_2+p_3)$, or the squares of the first three and in effect the three 
so-called angles derived from $p_1.p_2$, $p_1.p_3$ and $p_2.p_3$ or some other 
set. An explicit example of the dependence is provided in \cite{77,78} where 
the analytic expression for the one loop box integral is provided in four 
dimensions. There, by contrast, the lengths of the external momenta and two 
Mandelstam variables were used as the invariants. It was shown in \cite{77,78} 
that the planar box master with an arbitrary number of rungs is related to the 
master $3$-point planar triangle with the equivalent number of rungs. In 
practice that master integral is actually a function with three arguments where
each argument depends on combinations of the six underlying variables. Either 
way five dimensionless ratios plus one overall scale act as the independent 
variables for the renormalization of $4$-point functions at a subtraction 
point. For instance a specific example of this in QCD was provided in \cite{79}
where the quartic gluon vertex was studied at one loop at the fully symmetric 
point and latterly in \cite{80}. For a $4$-point function aside from the 
$6$-variable scheme there are in principle sub-variable schemes such as those
where one of the external momenta is initially nullified corresponding to an 
exceptional configuration to produce a $3$-variable scheme. As noted in 
\cite{46} provided one can carry out the renormalization in an infrared safe 
fashion using infrared rearrangement there would additionally be a set of 
$1$-variable schemes. The number basis in that instance should be the same as 
that of the $\MOMts$ schemes and in a similar way the $3$-variable schemes 
should involve the same suite of polylogarithm functions as those of an $x$ and
$y$ dependent MOM scheme. A hint of the appearance of more involved 
mathematical functions in schemes is already available from \cite{38} with the 
presence of an apparent non-multiple zeta number $P_{7,11}$ in the seven loop 
$\MSbar$ renormalization of $\phi^4$ theory. In discussing the potential 
ordering of schemes in this way we qualify the situation by mentioning that the
actual number of distinct $\half n(n-1)$-variable schemes for $n$-point 
functions is dependent on the field content of the underlying Lagrangian. As 
the QCD situation shows there are several $1$-variable schemes for each 
$3$-point vertex, as constructed earlier, and in a linear covariant gauge 
fixing there is only one $6$-variable scheme together with various lower 
variable schemes derived from it. This would complete the classification of all
possible massless kinematic based schemes in QCD. By contrast in $\phi^3$ 
theory there is only one $1$-variable and one $3$-variable scheme. The 
discussion of the $n$-variable scheme classification has rested on the vertex 
function. Within each particular $n$-variable scheme there is of course the 
further subdivision into the actual prescription to determine the 
renormalization constants themselves such as whether to include finite parts as
in the $\MOMts$ set or not in the $\MSbar$ case aside from a hybrid mixture 
akin to the $\RI$ scheme.

Returning to the theme of this article, which is the absence of $\zeta_4$ and
$\zeta_6$ in $\MOMtstars$ schemes, in light of the previous remarks it would 
seem that this property may be specifically confined to $1$-variable schemes. 
This is because the treatment of the $3$-point renormalization by one external 
momentum nullification immediately reduces those vertices to the $1$-variable 
case. From the higher variable scheme point of view if one has the general 
$3$-variable scheme with the finite part subtraction then the source of the 
$\zeta_4$ cancellation at four loops in the $\beta$-function could in principle
be investigated, say, in the limit to the $1$-variable case. At present the 
necessary three loop $3$-point master integrals are not known for non-zero $x$ 
and $y$; only the corresponding two loop masters are available, 
\cite{81,82,83,84,85}. If such a three loop arbitrary $x$ and $y$ 
renormalization could be carried out, it should be the case that taking the 
limit to the momentum configuration corresponding to one nullified external 
momentum produces an $\MOMtstars$ scheme as the endpoint. In that case the 
mathematical relations between the various types of polylogarithms, that ought 
to be the function basis for the three loop masters \cite{86}, may prove 
important in seeing how the $\zeta_4$ cancellation emerges in all the
renormalization group functions. Similar comments would equally apply to the 
next loop order to understand the relations that ensure the absence of 
$\zeta_6$. An additional observation based on the $1$-variable scheme situation
is that there will be parallel $\MOMts$ schemes for the $\half n(n-1)$-variable
schemes for $n$-point functions. In those cases it would be interesting to 
ascertain if there is an analogous set of functions that arise in the finite 
part of the Green's functions at a particular loop order but do not contribute 
to the renormalization group functions at the next loop in such an $\MOMts$ 
prescription.

Some of the points made in this section can be illustrated by an example. The
possibility that $\zeta_4$ was perturbatively absent in a situation which had 
physical significance was illuminated in \cite{44}. It centred on the Adler
$D$-function in the $\MSbar$ scheme. Subsequently this property was formulated
in a no-$\pi$ theorem in \cite{44}. Aside from the absence of $\zeta_4$ the 
theorem specified several conditions that involved what was termed 
$p$-integrals, \cite{44}. For the present discussion the relevant ones are that
for a massless correlator evaluated in a $\pi$-safe class, \cite{44}, using
$p$-integrals then it is $\pi$-free in a renormalization scheme that is free
of $\pi$, \cite{44}. Another way of expressing this is that the theorem applied
to massless correlation functions determined in $1$-variable schemes. It is 
straightforward to see that for $3$-variable schemes a different number basis 
structure is present. For instance if we define the perturbative expansion of 
the Adler $D$-function by
\begin{equation}
D(Q^2) ~=~ d_R C^{\mbox{\scriptsize{Adl}}}(a,\alpha)
\end{equation}
in the same notation as \cite{87}, where $d_R$~$=$~$3$ for $SU(3)$, then in the
$\MOMq$ scheme of \cite{51,52} for the same group we have 
\begin{eqnarray}
\left. C^{\mbox{\scriptsize{Adl}}}_{\MOMqss}(a,0) \right|^{SU(3)} &=& 1 + 4 a
+ \left[
- \frac{340}{81} \pi^2
- \frac{92}{9} \Nf
+ \frac{32}{3} \zeta_3 \Nf
+ \frac{170}{27} \pgone
+ \frac{463}{3}
- 176 \zeta_3
\right] a^2
\nonumber \\
&&
+ \left[
- \frac{1050461}{108} \zeta_3
- \frac{228443}{486} \pi^2
- \frac{53455}{2187} \pi^2 \pgone
- \frac{30722}{27} \Nf
\right. \nonumber \\
&& \left. ~~~~
- \frac{14960}{27} \zeta_3 \pgone
- \frac{12968}{243} \pgone \Nf
- \frac{5440}{243} \zeta_3 \pi^2 \Nf
- \frac{1600}{9} \zeta_5 \Nf
\right. \nonumber \\
&& \left. ~~~~
- \frac{64}{3} \zeta_3 \Nf^2
- \frac{16}{27} \pi^4 \Nf
+ \frac{2}{9} \pgthree \Nf
+ \frac{505}{1944} \pgthree
\right. \nonumber \\
&& \left. ~~~~
+ \frac{2720}{81} \zeta_3 \pgone \Nf
+ \frac{3032}{3} \zeta_3 \Nf
+ \frac{8800}{3} \zeta_5
+ \frac{25936}{729} \pi^2 \Nf
+ \frac{29920}{81} \zeta_3 \pi^2
\right. \nonumber \\
&& \left. ~~~~
+ \frac{48910}{6561} \pi^4
+ \frac{53455}{2916} \pgone^2
+ \frac{228443}{324} \pgone
+ 32 \Nf^2
+ 8679
\right] a^3
\nonumber \\
&&
+ \left[
- \frac{44274159239}{77760} \zeta_3
- \frac{15865609679}{349920} \pi^2
- \frac{14446746791}{104976} \zeta_5
\right. \nonumber \\
&& \left. ~~~~
+ \frac{15865609679}{233280} \pgone
- \frac{342716743}{1119744} \pgthree
- \frac{141906113}{14580} \pgone \Nf
\right. \nonumber \\
&& \left. ~~~~
- \frac{86603095}{23328} \pi^2 \pgone
- \frac{56330612}{10935} \zeta_3 \pi^2 \Nf
- \frac{39752681}{324} \Nf
\right. \nonumber \\
&& \left. ~~~~
- \frac{15012587}{52488} \pi^2 \pgone^2
- \frac{14454071}{196830} \pi^6
- \frac{6626701}{2834352} \pi^2 \pgthree
\right. \nonumber \\
&& \left. ~~~~
- \frac{3164155}{13122} \pi^4 \Nf
- \frac{1196852}{81} \zeta_3^2 \Nf
- \frac{981344}{243} \zeta_3 \Nf^2
- \frac{748000}{81} \zeta_5 \pi^2
\right. \nonumber \\
&& \left. ~~~~
- \frac{619520}{729} \zeta_3 \pi^4
- \frac{117040}{3} \zeta_7
- \frac{76621}{486} \pgone^2 \Nf
- \frac{68000}{81} \zeta_5 \pgone \Nf
\right. \nonumber \\
&& \left. ~~~~
- \frac{55330}{27} \zeta_3 \pgone^2
- \frac{46684}{2187} \pi^4 \pgone \Nf
- \frac{43130}{729} H_5
\right. \nonumber \\
&& \left. ~~~~
- \frac{40240}{243} \zeta_3 \pi^2 \pgone \Nf
- \frac{27520}{9} \zeta_5 \Nf^2
- \frac{16928}{81} \zeta_3 \pgone \Nf^2
\right. \nonumber \\
&& \left. ~~~~
- \frac{15248}{81} \pi^2 \Nf^2
- \frac{13246}{729} \zeta_3 \pgthree \Nf
- \frac{12473}{87480} \pgfive \Nf
\right. \nonumber \\
&& \left. ~~~~
- \frac{9376}{729} \pgone^3 \Nf
- \frac{5555}{243} \zeta_3 \pgthree
- \frac{4736}{729} H_5 \Nf
- \frac{3313}{2187} \pgthree \Nf^2
\right. \nonumber \\
&& \left. ~~~~
- \frac{1072}{9} \Nf^3
- \frac{256}{81} \zeta_3 \pi^4 \Nf^2
- \frac{164}{229635} H_6
- \frac{85}{81} \pi^2 \pgthree \Nf
\right. \nonumber \\
&& \left. ~~~~
- \frac{32}{27} \pi^2 \pgone \Nf^2
+ \frac{8}{9} \pgone^2 \Nf^2
+ \frac{32}{27} \zeta_3 \pgthree \Nf^2
\right. \nonumber \\
&& \left. ~~~~
+ \frac{85}{54} \pgone \pgthree \Nf
+ \frac{448}{9} \zeta_3 \Nf^3
+ \frac{640}{9} \zeta_5 \Nf^3
+ \frac{2560}{9} \zeta_3^2 \Nf^2
\right. \nonumber \\
&& \left. ~~~~
+ \frac{7624}{27} \pgone \Nf^2
+ \frac{10060}{81} \zeta_3 \pgone^2 \Nf
+ \frac{18752}{729} \pi^2 \pgone^2 \Nf
\right. \nonumber \\
&& \left. ~~~~
+ \frac{21280}{9} \zeta_7 \Nf
+ \frac{29096}{6561} \pi^4 \Nf^2
+ \frac{33856}{243} \zeta_3 \pi^2 \Nf^2
+ \frac{63707}{9} \Nf^2
\right. \nonumber \\
&& \left. ~~~~
+ \frac{136000}{243} \zeta_5 \pi^2 \Nf
+ \frac{153242}{729} \pi^2 \pgone \Nf
+ \frac{221320}{81} \zeta_3 \pi^2 \pgone
\right. \nonumber \\
&& \left. ~~~~
+ \frac{226688}{2187} \zeta_3 \pi^4 \Nf
+ \frac{374000}{27} \zeta_5 \pgone
+ \frac{433012}{32805} \pi^6 \Nf
\right. \nonumber \\
&& \left. ~~~~
+ \frac{2244703}{34992} \pgthree \Nf
+ \frac{4491806}{27} \zeta_3^2
+ \frac{6626701}{1889568} \pgone \pgthree
\right. \nonumber \\
&& \left. ~~~~
+ \frac{15012587}{104976} \pgone^3
+ \frac{15246509}{18895680} \pgfive
+ \frac{28165306}{3645} \zeta_3 \pgone \Nf
\right. \nonumber \\
&& \left. ~~~~
+ \frac{64243291}{354294} \pi^4 \pgone
+ \frac{86603095}{31104} \pgone^2
+ \frac{141906113}{21870} \pi^2 \Nf
\right. \nonumber \\
&& \left. ~~~~
+ \frac{214548299}{2430} \zeta_3 \Nf
+ \frac{257743792}{6561} \zeta_5 \Nf
+ \frac{274741727}{432}
+ \frac{862335313}{419904} \pi^4
\right. \nonumber \\
&& \left. ~~~~
+ \frac{1968292019}{43740} \zeta_3 \pi^2
- \frac{1968292019}{29160} \zeta_3 \pgone
\right] a^4 ~+~ O(a^5) 
\label{admomq}
\end{eqnarray}
where $H_5$ and $H_6$ are defined in the Supplemental Material of \cite{68} and
are shorthand for different combinations of harmonic or generalized 
polylogarithms. We have derived (\ref{admomq}) using the Landau gauge coupling 
constant mapping between the $\MSbar$ and $\MOMq$ schemes computed in 
\cite{51,52,67,68} which was applied to the $O(a^4)$ $\MSbar$ $D$-function of 
\cite{87}. As indicated earlier the $\MOMq$ scheme is a $3$-variable scheme but
in this case the two dimensionless variables $x$ and $y$ are both unity. What 
is evident in (\ref{admomq}) is that $\zeta_{2n}$~$\propto$~$\pi^{2n}$ appears 
in the $O(a^{n+1})$ term for $n$~$\geq$~$1$. This includes $\zeta_2$ which is 
absent in the $\MSbar$ and $\MOMts$ schemes. However (\ref{admomq}) does not 
violate the no-$\pi$ theorem because the master integrals underlying the MOM 
schemes are not $p$-integrals. In fact in this MOM configuration the 
$\zeta_{2n}$ contributions are each connected to $\psi^{(2n-1)}({\third})$ in 
the master integrals. More specifically if one makes the redefinitions to 
$\hat{\psi}^{(n)}({\third})$ via
\begin{eqnarray}
\pgone &=& \hat{\psi}^{(1)}({\third}) ~+~ 4 \zeta_2 ~~~,~~~
\pgthree ~=~ \hat{\psi}^{(3)}({\third}) ~+~ 240 \zeta_4 \nonumber \\
\pgfive &=& \hat{\psi}^{(5)}({\third}) ~+~ 43680 \zeta_6 
\label{psiredef}
\end{eqnarray}
then $\zeta_2$, $\zeta_4$ and $\zeta_6$ are effectively hidden but not absent. 
By this we mean that (\ref{psiredef}) is a shorthand for combinations that 
appear in the MOM external momentum configuration. It ought not to be 
interpreted as the same as the $\epsilon$ dependent redefinition of the zeta 
sequence in the same manner as that introduced in \cite{44} which is connected 
to the $\MOMts$ scheme exclusion of even zetas. 

There is one aspect worth highlighting in this example. The Adler $D$-function 
is constructed from the derivative of the $2$-point correlation function of the
vector current. As such it is a Green's function depending on one variable 
which is clearly the magnitude of the momentum transfer. However the coupling 
constant renormalization can be carried out in schemes other than those we have
designated as $1$-variable ones. For instance, the $\MOMq$ scheme is in the 
class of $3$-variable schemes but it clearly has $\pi^2$ contributions as well 
as derivatives of the Euler $\Gamma$-function. As this situation lies outside 
the conditions of the no-$\pi$ theorem there is no contradiction with it. 
Instead it merely illustrates how the situation changes with regard to the 
appearance of $\pi^2$ for the same physical quantity but in a different class 
of schemes. However what is not immediately clear concerns the $\zeta_{2n}$ 
dependence of a more general situation. The no-$\pi$ theorem makes no mention 
of whether the massless correlation function is restricted to that for two 
operators. For instance, in the case of a $3$-point gauge invariant operator 
correlation function the situation is more involved. This is assuming none of 
the three operators have a zero momentum flow which would reduce the 
correlation function to an effective $2$-point computation. For the $3$-point 
correlator the finite expression after renormalization should depend on a 
similar number basis, or its generalization for the non-unit values of the $x$ 
and $y$ variables, as that given in the $\beta$-functions of 
\cite{51,52,67,68}. It is therefore not clear if there is a scheme which would 
transform the finite part to the $\zeta_n$ and rational number basis of a 
$1$-variable scheme for a purely $3$-point operator correlation function. This 
may be the next task to study to understand the absence of $\zeta_{2n}$ in 
observable quantities.

One of the main reasons why we reviewed the renormalization group invariance of
critical exponents in the previous section is that it adds to the understanding
of the expression of the no-$\pi$ theorem. Recalling that the $\epsilon$ 
expansions of critical exponents at the Wilson-Fisher fixed point are scheme 
independent the $O(\epsilon^5)$ expansion of $\omega$ for $SU(3)$ in QCD is
\begin{eqnarray}
\left. \omega \right|^{SU(3)} &=&
\epsilon
+ 6 [19 \Nf - 153 ] \frac{\epsilon^2}{[2 \Nf - 33]^2} 
\nonumber \\
&&
+~ [ -~ 650 \Nf^3 + 14931 \Nf^2 - 233937 \Nf + 860139 ] 
\frac{\epsilon^3}{[2 \Nf - 33]^4}
\nonumber \\
&&
+~ [ -~ 8744 \Nf^5 
- 465984 \zeta_3 \Nf^4 
+ 264612 \Nf^4 
+ 16783200 \zeta_3 \Nf^3 
+ 14077854 \Nf^3 
\nonumber \\
&& ~~~~
- 194038416 \zeta_3 \Nf^2 
- 304090713 \Nf^2
+ 1068622632 \zeta_3 \Nf 
+ 2677244157 \Nf 
\nonumber \\
&& ~~~~
- 5658783768 \zeta_3 
- 6752933307 ] \frac{\epsilon^4}{6 [2 \Nf - 33]^6}
\nonumber \\
&&
+~ [ 175104 \zeta_3 \Nf^7 
- 38560 \Nf^7 
+ 10041600 \zeta_3 \Nf^6 
- 5591808 \zeta_4 \Nf^6 
- 4769280 \zeta_5 \Nf^6 
\nonumber \\
&& ~~~~
+ 13696280 \Nf^6 
- 1091992320 \zeta_3 \Nf^5 
+ 385928064 \zeta_4 \Nf^5 
+ 675866880 \zeta_5 \Nf^5 
\nonumber \\
&& ~~~~
- 804202392 \Nf^5 
+ 38019806208 \zeta_3 \Nf^4 
- 10496977920 \zeta_4 \Nf^4 
- 30361478400 \zeta_5 \Nf^4 
\nonumber \\
&& ~~~~
+ 17650466742 \Nf^4
- 694400454720 \zeta_3 \Nf^3 
+ 144493398720 \zeta_4 \Nf^3 
\nonumber \\
&& ~~~~
+ 639988905600 \zeta_5 \Nf^3 
- 280199346390 \Nf^3 
+ 6778342959696 \zeta_3 \Nf^2 
\nonumber \\
&& ~~~~
- 1125003472560 \zeta_4 \Nf^2 
- 7142270968800 \zeta_5 \Nf^2 
+ 2651536463832 \Nf^2 
\nonumber \\
&& ~~~~
- 33418251568944 \zeta_3 \Nf 
+ 5732068510872 \zeta_4 \Nf 
+ 43054184851920 \zeta_5 \Nf 
\nonumber \\
&& ~~~~
- 13351743621324 \Nf
+ 81853049696616 \zeta_3 
- 18487246570056 \zeta_4 
\nonumber \\
&& ~~~~
- 120758445609120 \zeta_5
+ 24360811371837 ] \frac{\epsilon^5}{12 [2 \Nf - 33]^8} ~+~
O(\epsilon^6)
\label{omegasu3}
\end{eqnarray}
with the exponents for the the gluon, ghost, quark and quark mass taking a
similar form. Clearly (\ref{omegasu3}) depends on $\zeta_4$ at $O(\epsilon^5)$ 
as do the other exponents. Endeavouring to remove such a contribution is not 
possible as (\ref{omegasu3}) is independent of the scheme. Indeed we computed 
$\omega$ directly for each of the $\MOMtstars$ scheme and verified that the 
same expression as (\ref{omegasu3}) resulted. So for example the $\epsilon$ 
dependent mapping of the zetas of \cite{44} cannot be applied. That in effect 
is related to a scheme change and such a change has already been incorporated
within the $\MOMtstars$ construction. Moreover the presence of $\zeta_4$ does 
not contradict the criteria of the no-$\pi$ theorem of \cite{44}. While the 
computation that leads to (\ref{omegasu3}) was also carried out in the $\MSbar$
scheme using $p$-integrals it can equally well be carried out in any of the MOM
schemes of \cite{51,52} which do not use $p$-integrals. The same result is 
obtained. 

As a final comment on (\ref{omegasu3}) one can isolate the $\zeta_4$ 
contribution at $O(\epsilon^5)$ for an arbitrary colour group. Denoting this by
$\left. \omega \right|^{\zeta_4}_{\epsilon^5}$ we have
\begin{eqnarray}
\left. \omega \right|^{\zeta_4}_{\epsilon^5} &=& 
[ 11 C_A^4 \NA 
- 102 C_A^3 \NA \Nf T_F 
+ 164 C_A^2 C_F \NA \Nf T_F
- 56 C_A^2 \NA \Nf^2 T_F^2 
\nonumber \\
&& ~
- 88 C_A C_F^2 \NA \Nf T_F 
- 112 C_A C_F \NA \Nf^2 T_F^2 
+ 176 C_F^2 \NA \Nf^2 T_F^2
- 528 d_A^{abcd} d_A^{abcd} 
\nonumber \\
&& ~
+ 1248 d_A^{abcd} d_F^{abcd} \Nf
- 384 d_F^{abcd} d_F^{abcd} \Nf^2 ] 
\frac{162 \zeta_4}{[ 11 C_A - 4 \Nf T_F ]^4 \NA} ~. 
\label{omegaz4ep5}
\end{eqnarray}
We noted that in $SU(3)$ the $O(a^5)$ coefficient of $\zeta_4$ in the coupling 
constant maps of (\ref{ccmaps}) was the same in all the $\MOMtstars$ schemes
when $\Nf$~$=$~$3$. Aside from $\zeta_4$ there was no commonality for the odd
zetas across all the scheme maps. This allows us to track down some aspects of
how the $\zeta_4$ contribution in the $\MSbar$ $\beta$-function at five loops 
leads to (\ref{omegaz4ep5}) as well as how the non-appearance of $\zeta_4$ in 
$\beta_{\MOMtstarss}(a,\alpha)$ still results in its presence in this scheme 
independent exponent. It transpires that the route to (\ref{omegaz4ep5}) for 
these distinct schemes is different. First we isolated the $\zeta_4$ 
coefficient at $O(a^5)$ of (\ref{ccmaps}) for a general colour group in the 
Landau gauge and found that it is {\em precisely} proportional to the numerator
of (\ref{omegaz4ep5}) for all $\Nf$ and $\MOMtstars$ schemes. The constant of 
proportionality can be accounted for by the one loop coefficient of the 
$\beta$-function which is scheme independent. That term in the mappings can be 
traced back to the $O(\epsilon)$ contribution to the {\em four} loop term of 
$\beta_{\MOMtstarss}(a,\alpha)$. That means it is present in the finite part
of the vertex function for bare variables. This can be verified by noting that
the coefficient of $\zeta_4$ at five loops in the $\MSbar$ $\beta$-function is
proportional to the numerator of (\ref{omegaz4ep5}). The actual coefficient is
the product of the numerator and the one loop coefficient of the
$\beta$-function as well as a rational. In other words within the derivation of
(\ref{omegaz4ep5}) the $\zeta_4$ contribution emerges via two different routes 
depending on which scheme is used. In the $\MSbar$ case it appears in $\omega$
directly from the five loop term in the $\beta$-function. By constrast it is
absent in the $\MOMtstars$ $\beta$-function in four dimensions but present for
non-zero $\epsilon$ at four loops. In carrying out this analysis what we are 
basically summarizing is the same process that the $\epsilon$ dependent 
redefinition of the zeta series of \cite{44,45} is effecting but using the 
scheme independent $\omega$ as the pivot point to trace the details. We close 
the section by remarking that with respect to the classification introduced 
here one could regard $\omega$ and other such exponents as being determined in 
a $0$-variable scheme. This is partly because exponents are dimensionless 
quantities as there is no scale at a critical point. In turn this means that 
there is no underlying single momentum invariant similar to that which is 
present in the $1$-variable ones.

\sect{Discussion.}

One of the main aims of this study was to ascertain whether the extension of
the so-called $\MOMts$ schemes that were examined in earlier articles at lower
loop order retained the property of having no explicit $\pi^2$ terms in the
renormalization group functions at five loops in QCD for all values of $\alpha$
in a linear covariant gauge fixing. By exploiting the available {\sc Forcer} 
data on the bare $2$- and $3$-point functions, \cite{63}, we were indeed able 
to demonstrate that this is the case for the various single scale $3$-point 
vertices of (\ref{greenmomdef}). Moreover this observation in one sense both 
confirms and extends the study of \cite{46} which centred on what was termed 
there as AD theories. These are ones which have symmetries, such as gauge 
symmetry or supersymmetry, that means the coupling renormalization constant is 
determined by a Ward identity that relates it to the renormalization of the 
fields. In QCD one such AD scheme was already known about which was the $\mMOM$
scheme based on Taylor's theorem that the ghost-gluon vertex function is finite
in the Landau gauge. So the renormalization of the coupling is constructed 
purely from the ghost and gluon renormalization constants. From the available 
five loop renormalization group functions, \cite{43,63,64}, this is implicitly 
evident but only in the Landau gauge as can be verified by examining the 
$\alpha$~$=$~$0$ $\MOMtccgzcs$ scheme expressions in Appendix B. This 
observation in effect became a focal point for realizing that if the defining 
$\mMOM$ constraint on the coupling constant renormalization was removed then 
the absence of $\pi^2$ in the $\MOMtccgzcs$ scheme and the remaining 
$\MOMtstars$ schemes should follow for all $\alpha$. In other words it should 
be possible to extend the groundwork analysis of \cite{46} to non-AD theories. 
In one way supportive of that possibility is the fact that in QCD the gauge 
parameter could be regarded as a second coupling and \cite{46} examined the 
$\MOMts$ construction in a multicoupling theory indicating that the $\zeta_4$ 
cancellation would persist to six loops in some theories. Although what we have
examined here has to be qualified by noting that a perturbative expansion is 
not carried out in the gauge parameter itself.

To construct an all orders proof of the $\zeta_{2n}$ absence may not be 
straightforward and the Hopf algebra approach of \cite{58} in the Wess-Zumino 
model, motivated by the earlier work of \cite{88,89}, might allow for deeper 
understanding. For instance one feature of the $\MOMts$ scheme that seems to 
lie at its heart is that in the defining prescription the $3$-point functions 
are quotiented by the $2$-point functions of the relevant fields of that vertex
prior to the remainder being removed from the $3$-point vertex. This tallies 
with the approach of \cite{44,45}. In graphs that involve a simple pole in 
$\epsilon$ and a residue that depends on $\zeta_3$ the finite part will contain
$\zeta_4$ with its predetermined coefficient, \cite{44,45}. The removal of the 
finite part in a $\MOMts$ prescription means that it will contribute at the 
next order via the counterterms and thereby affect the coefficient of 
$\zeta_4$. Indeed we were able to verify this in dissecting the passage of 
$\zeta_4$ from the $\epsilon$-dependent $\beta$-functions to how it appears in 
a scheme independent quantity. We need to qualify the absence of $\pi^2$ 
dependence at any higher loop order in a renormalization group function by 
noting that one would also have to prove that there are no single scale master 
Feynman graphs whose leading term in its $\epsilon$ expansion is proportional 
to $\pi^2$. From the high order four dimensional examples that are already 
available in, for instance \cite{38,90,91}, no such cases appear to be known. 

Perhaps one place to examine these ideas in further detail would be in other 
spacetime dimensions. For instance, $\phi^3$ theory in six dimensions has 
already been noted as a potential laboratory to study $n$-variable schemes. As 
alluded to earlier examining $\phi^3$ theory in principle would first require 
the construction and analysis of the $3$-point Schwinger-Dyson equation. In the 
single field case the coupling constant renormalization is related to the 
renormalization of the mass of the $\phi$ field equating it to a $2$-point
function renormalization. Such a relationship however does not extend to the 
cubic theory with a symmetry. Another instance of where using a $\MOMts$ scheme 
might produce an interesting structure for a $\beta$-function is the two 
dimensional nonlinear $\sigma$ model with ${\cal N}$~$=$~$1$ supersymmetry. We 
note that the location of $\zeta_n$ in the $\beta$-function of these two 
dimensional supersymmetric models is that $\zeta_{L-1}$ appears for the first 
time at $L$ loops starting with $L$~$=$~$4$. Such supersymmetric models have 
been renormalized to four loops in the $\MSbar$ scheme on a general manifold 
\cite{92,93,94} and revealed that there are no contributions after the one loop
one until four loops. At that order the coefficient of the $\beta$-function 
involves only $\zeta_3$. For nonlinear $\sigma$ models with 
${\cal N}$~$=$~$2$ supersymmetry the four loop term has the same property. The
situation beyond four loops has been probed in several ways. Restricting the 
geometry to the $N$ dimensional sphere one can compute the $1/N$ corrections 
to the $\beta$-function in the large $N$ expansion, \cite{95,96}. Aside from 
being an independent confirmation of the $\beta$-function of the general 
geometry of \cite{92,93}, the result indicated that the five loop term would 
involve $\zeta_4$ in the $\MSbar$ scheme. In the ${\cal N}$~$=$~$2$ 
supersymmetric case the $\beta$-function was examined at five loops explicitly 
for K\"{a}hler manifolds in \cite{97}. Interestingly in that general geometry 
the five loop term can be made to vanish by a particular scheme choice 
\cite{97}. Whether that scheme has a connection with the $\MOMts$ prescription 
would be interesting to ascertain. If so it would tally with the absence of 
$\zeta_4$ terms in that set of renormalization prescriptions. At six loop order
it was argued in \cite{98} that there would be a non-zero contribution, solely 
involving $\zeta_5$, which remained even after any field redefinition. In the 
case of the $N$-sphere the five, six and seven loop coefficients depended only 
on zetas, \cite{95}. By this we specifically mean that $\zeta_4$ and the pairs 
$\{\zeta_3,\zeta_5\}$ and $\{\zeta_3^2,\zeta_6\}$ appear respectively at 
$O(1/N^2)$. Other $\zeta_n$ contributions could of course arise at higher 
orders in $1/N$. The field anomalous dimension had a similar structure, 
\cite{96,99}. 

Given these observations it might be of interest to see whether the five loop 
$\zeta_4$ and seven loop $\zeta_6$ contributions are absent in a direct 
$\MOMts$ scheme computation. The work of \cite{97} suggests this might be the 
case for the former. Equally it would be interesting to see how a scheme
transformation similar to the one for ${\cal N}$~$=$~$2$ supersymmetry 
discussed in \cite{97} would affect the zeta structure of an ${\cal N}$~$=$~$1$
theory. For instance from the results in the ${\cal N}$~$=$~$2$ $\sigma$ model 
it appears that only zetas appear at $L$ loops whose weight is $(L-1)$ for 
$L$~$\geq$~$4$. At seven loops in the ${\cal N}$~$=$~$1$ case it is known that 
$\zeta_3^2$ and $\zeta_6$ are present, \cite{95}. Therefore for a 
transformation to a $\MOMts$ scheme similar to that discussed here, the 
$\zeta_3^2$ contribution for the $N$-sphere should remain but $\zeta_6$ ought 
to be absent. In the ${\cal N}$~$=$~$2$ case the latter ought also to be absent
but the status of $\zeta_3^2$ in the $\beta$-function after a transformation is
not clear. While renormalizing these two dimensional supersymmetric $\sigma$ 
models is highly non-trivial beyond one loop, \cite{92,93,94,98}, and accessing
the various higher large $N$ orders is equally computationally demanding, the 
coefficients of the $\beta$-function would appear to be only zetas and multiple
zetas with no rationals, \cite{96,97,99}. It would therefore seem that these 
supersymmetric models might offer a future testing ground for analysing even 
zeta cancellation in the $\MOMts$ prescription at a deeper order beyond the 
five loop one considered here. However, the situation with regard to 
supersymmetric gauge theories is more intricate from the point of view of the
$C$-scheme concept. For instance prior to \cite{59,60} the exact NSVZ
(Novikov-Shifman-Vainshtein-Zakharov) $\beta$-function, \cite{100,101,102}, was
examined in \cite{103}. That $\beta$-function has similarities with the
$C$-scheme $\beta$-function of \cite{59,60}. Indeed a comparison between the 
${\cal N}$~$=$~$0$ and $1$ $\beta$-functions in gauge theories was carried out 
in \cite{104}. However as yet a detailed analysis of the $\zeta_{2n}$ 
dependence of all the renormalization group functions of supersymmetric gauge 
theories at high loop order has not been considered to the same depth as the 
non-supersymmetic ones.

Finally, having established the $\MOMtstars$ schemes have the property that 
there are no even zetas to five loops in QCD there is scope now to apply these 
schemes to explore what effect they have on phenomenological precision and
whether they can be employed for estimating theory errors. For instance the 
$C$-scheme was used, \cite{59}, to study $e^+ e^-$ scattering and $\tau$ decays
into hadrons. It was suggested that this scheme produces a scheme invariant 
scale running. Therefore it would seem appropriate to employ the $\MOMtstars$ 
data now to complement that study to ascertain what effect the absence of 
$\zeta_4$ and $\zeta_6$ has and to see if there is a similar reduction in scale
dependence. Equally the other question of what structures are absent in the 
analogous concept of $\MOMts$ scheme renormalization of $n$-variable schemes 
would be an interesting avenue to pursue. 

\vspace{1cm}
\noindent
{\bf Acknowledgements.} This work was carried out with the support of the STFC
Consolidated Grant ST/T000988/1. For the purpose of open access, the author
has applied a Creative Commons Attribution (CC-BY) licence to any Author
Accepted Manuscript version arising. The data representing the main results
here are accessible in electronic form from the arXiv ancillary directory
associated with the article. The author gratefully appreciates discussions with
I. Jack and R.H. Mason.

\appendix

\sect{$SU(3)$ Landau gauge $\beta$-functions.}

In this Appendix we record each of the $\MOMtstars$ $\beta$-functions in the 
$SU(3)$ colour group for the Landau gauge in order to clearly show that the
only $\zeta_n$ dependence is $\zeta_3$, $\zeta_5$ and $\zeta_7$ to five loops. 
First for the two ghost-gluon vertex schemes we have 
\begin{eqnarray}
\beta^{SU(3)}_{\MOMtccgzcss}(a,0) &=&
\left[
\frac{2}{3} \Nf
- 11
\right] a^2
+ \left[
\frac{38}{3} \Nf
- 102
\right] a^3
\nonumber \\
&&
+ \left[
\frac{3861}{8} \zeta_3
- \frac{28965}{8}
- \frac{989}{54} \Nf^2
- \frac{175}{12} \zeta_3 \Nf
- \frac{8}{9} \zeta_3 \Nf^2
+ \frac{7715}{12} \Nf
\right] a^4
\nonumber \\
&&
+ \left[
- \frac{1380469}{8}
- \frac{1027375}{144} \zeta_5 \Nf
- \frac{736541}{324} \Nf^2
- \frac{516881}{72} \zeta_3 \Nf
- \frac{16}{9} \zeta_3 \Nf^3
\right. \nonumber \\
&& \left. ~~~~
+ \frac{800}{27} \Nf^3
+ \frac{6547}{27} \zeta_3 \Nf^2
+ \frac{9280}{27} \zeta_5 \Nf^2
+ \frac{625317}{16} \zeta_3
+ \frac{772695}{32} \zeta_5
\right. \nonumber \\
&& \left. ~~~~
+ \frac{970819}{24} \Nf
\right] a^5
\nonumber \\
&&
+ \left[
- \frac{21619456551}{4096} \zeta_7
- \frac{18219328375}{6912} \zeta_5 \Nf
- \frac{10327103555}{20736} \zeta_3 \Nf
\right. \nonumber \\
&& \left. ~~~~
- \frac{3248220045}{256}
+ \frac{4922799165}{512} \zeta_5
+ \frac{24870449471}{18432} \zeta_7 \Nf
\right. \nonumber \\
&& \left. ~~~~
+ \frac{115659378547}{31104} \Nf
- \frac{833934985}{2592} \Nf^2
- \frac{26952037}{432} \zeta_7 \Nf^2
\right. \nonumber \\
&& \left. ~~~~
- \frac{299875}{54} \zeta_5 \Nf^3
- \frac{82869}{32} \zeta_3^2 \Nf
- \frac{59531}{36} \zeta_3^2 \Nf^2
- \frac{2617}{27} \Nf^4
\right. \nonumber \\
&& \left. ~~~~
- \frac{304}{27} \zeta_3 \Nf^4
+ \frac{1760}{27} \zeta_5 \Nf^4
+ \frac{2240}{27} \zeta_3^2 \Nf^3
+ \frac{129869}{162} \zeta_3 \Nf^3
\right. \nonumber \\
&& \left. ~~~~
+ \frac{3249767}{324} \Nf^3
+ \frac{7696161}{64} \zeta_3^2
+ \frac{13019053}{1296} \zeta_3 \Nf^2
+ \frac{65264845}{324} \zeta_5 \Nf^2
\right. \nonumber \\
&& \left. ~~~~
+ \frac{1064190195}{512} \zeta_3
\right] a^6 ~+~ O(a^7) 
\end{eqnarray}
and
\begin{eqnarray}
\beta^{SU(3)}_{\MOMtccgzgss}(a,0) &=&
\left[
\frac{2}{3} \Nf
- 11
\right] a^2
+ \left[
\frac{38}{3} \Nf
- 102
\right] a^3
\nonumber \\
&&
+ \left[
- \frac{28263}{8}
- \frac{535}{27} \Nf^2
- \frac{175}{12} \zeta_3 \Nf
- \frac{8}{9} \zeta_3 \Nf^2
+ \frac{3799}{6} \Nf
+ \frac{3861}{8} \zeta_3
\right] a^4
\nonumber \\
&&
+ \left[
- \frac{5516125}{32}
- \frac{1039525}{144} \zeta_5 \Nf
- \frac{198992}{81} \Nf^2
- \frac{64411}{9} \zeta_3 \Nf
- \frac{16}{9} \zeta_3 \Nf^3
\right. \nonumber \\
&& \left. ~~~~
+ \frac{980}{27} \Nf^3
+ \frac{9280}{27} \zeta_5 \Nf^2
+ \frac{13445}{54} \zeta_3 \Nf^2
+ \frac{295581}{8} \zeta_3
+ \frac{817245}{32} \zeta_5
\right. \nonumber \\
&& \left. ~~~~
+ \frac{1953167}{48} \Nf
\right] a^5
\nonumber \\
&&
+ \left[
- \frac{67000185565}{27648} \zeta_5 \Nf
- \frac{25922636709}{2048}
- \frac{22470285835}{41472} \zeta_3 \Nf
\right. \nonumber \\
&& \left. ~~~~
- \frac{18877588191}{4096} \zeta_7
- \frac{3503317141}{10368} \Nf^2
+ \frac{2815217703}{1024} \zeta_3
\right. \nonumber \\
&& \left. ~~~~
+ \frac{16486752015}{2048} \zeta_5
+ \frac{23141911415}{18432} \zeta_7 \Nf
+ \frac{929972881523}{248832} \Nf
\right. \nonumber \\
&& \left. ~~~~
- \frac{12779459}{216} \zeta_7 \Nf^2
- \frac{612845}{108} \zeta_5 \Nf^3
- \frac{267689}{144} \zeta_3^2 \Nf^2
- \frac{3022}{27} \Nf^4
\right. \nonumber \\
&& \left. ~~~~
- \frac{304}{27} \zeta_3 \Nf^4
+ \frac{1760}{27} \zeta_5 \Nf^4
+ \frac{2240}{27} \zeta_3^2 \Nf^3
+ \frac{491045}{648} \zeta_3 \Nf^3
+ \frac{1066263}{256} \zeta_3^2 \Nf
\right. \nonumber \\
&& \left. ~~~~
+ \frac{28965085}{2592} \Nf^3
+ \frac{33124113}{512} \zeta_3^2
+ \frac{62058733}{5184} \zeta_3 \Nf^2
+ \frac{1017487675}{5184} \zeta_5 \Nf^2
\right] a^6 
\nonumber \\
&&
+~ O(a^7) ~.
\end{eqnarray}
The other scheme based on the triple gluon vertex produces
\begin{eqnarray}
\beta^{SU(3)}_{\MOMtgggzggss}(a,0) &=&
\left[
\frac{2}{3} \Nf
- 11
\right] a^2
+ \left[
\frac{38}{3} \Nf
- 102
\right] a^3
\nonumber \\
&&
+ \left[
- \frac{186747}{64}
- \frac{65}{6} \zeta_3 \Nf
- \frac{8}{9} \Nf^3
- \frac{8}{9} \zeta_3 \Nf^2
+ \frac{829}{54} \Nf^2
+ \frac{1683}{4} \zeta_3
+ \frac{35473}{96} \Nf
\right] a^4
\nonumber \\
&&
+ \left[
- \frac{20783939}{128}
- \frac{1464379}{648} \Nf^2
- \frac{1323259}{144} \zeta_3 \Nf
- \frac{908995}{144} \zeta_5 \Nf
- \frac{320}{81} \Nf^4
\right. \nonumber \\
&& \left. ~~~~
- \frac{64}{9} \zeta_3 \Nf^3
+ \frac{3164}{27} \Nf^3
+ \frac{7540}{27} \zeta_5 \Nf^2
+ \frac{12058}{27} \zeta_3 \Nf^2
+ \frac{900075}{32} \zeta_5
\right. \nonumber \\
&& \left. ~~~~
+ \frac{1300563}{32} \zeta_3
+ \frac{2410799}{64} \Nf
\right] a^5
\nonumber \\
&&
+ \left[
- \frac{46418845041}{4096}
- \frac{26430396425}{13824} \zeta_5 \Nf
- \frac{4685253111}{4096} \zeta_7
\right. \nonumber \\
&& \left. ~~~~
+ \frac{2800824887}{18432} \zeta_7 \Nf
+ \frac{7708557555}{1024} \zeta_5
+ \frac{1422229465579}{497664} \Nf
\right. \nonumber \\
&& \left. ~~~~
- \frac{1924634167}{10368} \zeta_3 \Nf
- \frac{421916191}{2592} \Nf^2
- \frac{54502201}{2592} \zeta_3 \Nf^2
\right. \nonumber \\
&& \left. ~~~~
- \frac{36167831}{3456} \zeta_7 \Nf^2
- \frac{29633175}{512} \zeta_3^2
- \frac{433589}{648} \Nf^3
- \frac{373877}{144} \zeta_3^2 \Nf^2
\right. \nonumber \\
&& \left. ~~~~
- \frac{313255}{54} \zeta_5 \Nf^3
- \frac{832}{27} \zeta_3 \Nf^4
- \frac{416}{27} \Nf^5
+ \frac{998}{27} \zeta_3^2 \Nf^3
+ \frac{1323}{4} \zeta_7 \Nf^3
\right. \nonumber \\
&& \left. ~~~~
+ \frac{1760}{27} \zeta_5 \Nf^4
+ \frac{33547}{81} \Nf^4
+ \frac{645595}{324} \zeta_3 \Nf^3
+ \frac{9288863}{256} \zeta_3^2 \Nf
\right. \nonumber \\
&& \left. ~~~~
+ \frac{196487181}{256} \zeta_3
+ \frac{437857925}{2592} \zeta_5 \Nf^2
\right] a^6 ~+~ O(a^7) ~.
\end{eqnarray}
Finally the schemes derived from the quark-gluon vertex lead to
\begin{eqnarray}
\beta^{SU(3)}_{\MOMtqqgzgss}(a,0) &=&
\left[
\frac{2}{3} \Nf
- 11
\right] a^2
+ \left[
+ \frac{38}{3} \Nf
- 102
\right] a^3
\nonumber \\
&&
+ \left[
- \frac{150931}{48}
- \frac{797}{54} \Nf^2
- \frac{118}{3} \zeta_3 \Nf
- \frac{8}{9} \zeta_3 \Nf^2
+ \frac{42089}{72} \Nf
+ 891 \zeta_3
\right] a^4
\nonumber \\
&&
+ \left[
- \frac{44627671}{576}
- \frac{3718495}{96} \zeta_5
- \frac{959761}{648} \Nf^2
- \frac{110887}{12} \zeta_3 \Nf
- \frac{16}{9} \zeta_3 \Nf^3
\right. \nonumber \\
&& \left. ~~~~
+ \frac{164}{9} \Nf^3
+ \frac{2800}{27} \zeta_5 \Nf^2
+ \frac{7771}{27} \zeta_3 \Nf^2
+ \frac{91645}{144} \zeta_5 \Nf
+ \frac{1458193}{24} \zeta_3
\right. \nonumber \\
&& \left. ~~~~
+ \frac{22417595}{864} \Nf
\right] a^5
\nonumber \\
&&
+ \left[
- \frac{110207309485}{27648}
- \frac{37281587675}{20736} \zeta_5 \Nf
+ \frac{6054470821}{9216} \zeta_7 \Nf
\right. \nonumber \\
&& \left. ~~~~
+ \frac{55672020805}{13824} \zeta_5
+ \frac{244350005119}{124416} \Nf
- \frac{2027774803}{10368} \Nf^2
\right. \nonumber \\
&& \left. ~~~~
- \frac{1790283341}{2048} \zeta_7
- \frac{249619937}{432} \zeta_3
- \frac{63316501}{2592} \zeta_3 \Nf^2
- \frac{31626203}{864} \zeta_7 \Nf^2
\right. \nonumber \\
&& \left. ~~~~
- \frac{5338625}{64} \zeta_3^2 \Nf
- \frac{159320}{27} \zeta_5 \Nf^3
- \frac{1657}{27} \Nf^4
- \frac{1648}{27} \zeta_3^2 \Nf^3
- \frac{304}{27} \zeta_3 \Nf^4
\right. \nonumber \\
&& \left. ~~~~
+ \frac{1760}{27} \zeta_5 \Nf^4
+ \frac{127765}{81} \zeta_3 \Nf^3
+ \frac{172645}{36} \zeta_3^2 \Nf^2
+ \frac{230059}{36} \Nf^3
\right. \nonumber \\
&& \left. ~~~~
+ \frac{13628215}{1728} \zeta_3 \Nf
+ \frac{44150073}{128} \zeta_3^2
+ \frac{458425445}{2592} \zeta_5 \Nf^2
\right] a^6 ~+~ O(a^7) \nonumber \\
\beta^{SU(3)}_{\MOMtqqgzgTss}(a,0) &=&
\left[
\frac{2}{3} \Nf
- 11
\right] a^2
+ \left[
\frac{38}{3} \Nf
- 102
\right] a^3
\nonumber \\
&&
+ \left[
- \frac{185039}{48}
- \frac{953}{54} \Nf^2
- \frac{8}{9} \zeta_3 \Nf^2
+ \frac{47221}{72} \Nf
- 48 \zeta_3 \Nf
+ 1034 \zeta_3
\right] a^4
\nonumber \\
&&
+ \left[
- \frac{32456317}{192}
- \frac{3369385}{432} \zeta_5 \Nf
- \frac{1412065}{648} \Nf^2
- \frac{851009}{108} \zeta_3 \Nf
- \frac{16}{9} \zeta_3 \Nf^3
\right. \nonumber \\
&& \left. ~~~~
+ \frac{740}{27} \Nf^3
+ \frac{5875}{27} \zeta_3 \Nf^2
+ \frac{11440}{27} \zeta_5 \Nf^2
+ \frac{1269361}{32} \Nf
+ \frac{3841475}{288} \zeta_5
\right. \nonumber \\
&& \left. ~~~~
+ \frac{4134361}{72} \zeta_3
\right] a^5
\nonumber \\
&&
+ \left[
- \frac{118778905711}{9216}
- \frac{83114846969}{18432} \zeta_7
- \frac{19658419625}{6912} \zeta_5 \Nf
\right. \nonumber \\
&& \left. ~~~~
- \frac{3287597531}{10368} \Nf^2
+ \frac{37311626107}{27648} \zeta_7 \Nf
+ \frac{43800203815}{4608} \zeta_5
\right. \nonumber \\
&& \left. ~~~~
+ \frac{463743395407}{124416} \Nf
- \frac{827828681}{1296} \zeta_3 \Nf
- \frac{139465865}{384} \zeta_3^2
\right. \nonumber \\
&& \left. ~~~~
- \frac{56355551}{864} \zeta_7 \Nf^2
- \frac{158060}{27} \zeta_5 \Nf^3
- \frac{52295}{12} \zeta_3^2 \Nf^2
- \frac{2437}{27} \Nf^4
- \frac{304}{27} \zeta_3 \Nf^4
\right. \nonumber \\
&& \left. ~~~~
+ \frac{1760}{27} \zeta_5 \Nf^4
+ \frac{3536}{27} \zeta_3^2 \Nf^3
+ \frac{55165}{81} \zeta_3 \Nf^3
+ \frac{782282}{81} \Nf^3
\right. \nonumber \\
&& \left. ~~~~
+ \frac{33559267}{576} \zeta_3^2 \Nf
+ \frac{41202899}{2592} \zeta_3 \Nf^2
+ \frac{570243785}{2592} \zeta_5 \Nf^2
\right. \nonumber \\
&& \left. ~~~~
+ \frac{1248094123}{384} \zeta_3
\right] a^6 ~+~ O(a^7) \nonumber \\
\beta^{SU(3)}_{\MOMtqqgzqss}(a,0) &=&
\left[
\frac{2}{3} \Nf
- 11
\right] a^2
+ \left[
\frac{38}{3} \Nf
- 102
\right] a^3
\nonumber \\
&&
+ \left[
- \frac{29559}{8}
- \frac{989}{54} \Nf^2
- \frac{64}{3} \zeta_3 \Nf
- \frac{8}{9} \zeta_3 \Nf^2
+ \frac{7769}{12} \Nf
+ 594 \zeta_3
\right] a^4
\nonumber \\
&&
+ \left[
- \frac{2795027}{16}
- \frac{1016935}{144} \zeta_5 \Nf
- \frac{737837}{324} \Nf^2
- \frac{67939}{9} \zeta_3 \Nf
- \frac{16}{9} \zeta_3 \Nf^3
\right. \nonumber \\
&& \left. ~~~~
+ \frac{800}{27} \Nf^3
+ \frac{6709}{27} \zeta_3 \Nf^2
+ \frac{9280}{27} \zeta_5 \Nf^2
+ \frac{174207}{4} \zeta_3
+ \frac{487751}{12} \Nf
\right. \nonumber \\
&& \left. ~~~~
+ \frac{734415}{32} \zeta_5
\right] a^5
\nonumber \\
&&
+ \left[
- \frac{18215306455}{6912} \zeta_5 \Nf
- \frac{10920152021}{20736} \zeta_3 \Nf
- \frac{9939037647}{2048} \zeta_7
\right. \nonumber \\
&& \left. ~~~~
+ \frac{4803715005}{512} \zeta_5
+ \frac{11859668359}{9216} \zeta_7 \Nf
+ \frac{58653420029}{15552} \Nf
\right. \nonumber \\
&& \left. ~~~~
- \frac{421368587}{1296} \Nf^2
- \frac{51878571}{4}
- \frac{25991539}{432} \zeta_7 \Nf^2
- \frac{152570}{27} \zeta_5 \Nf^3
\right. \nonumber \\
&& \left. ~~~~
- \frac{15776}{9} \zeta_3^2 \Nf^2
- \frac{2617}{27} \Nf^4
- \frac{304}{27} \zeta_3 \Nf^4
+ \frac{1760}{27} \zeta_5 \Nf^4
+ \frac{2240}{27} \zeta_3^2 \Nf^3
\right. \nonumber \\
&& \left. ~~~~
+ \frac{127439}{162} \zeta_3 \Nf^3
+ \frac{495753}{128} \zeta_3^2 \Nf
+ \frac{3273095}{324} \Nf^3
+ \frac{10403415}{256} \zeta_3^2
\right. \nonumber \\
&& \left. ~~~~
+ \frac{28286855}{2592} \zeta_3 \Nf^2
+ \frac{66058375}{324} \zeta_5 \Nf^2
+ \frac{1222682025}{512} \zeta_3
\right] a^6 ~+~ O(a^7) 
\end{eqnarray}
and 
\begin{eqnarray}
\beta^{SU(3)}_{\MOMtqqgzqTss}(a,0) &=&
\left[
\frac{2}{3} \Nf
- 11
\right] a^2
+ \left[
\frac{38}{3} \Nf
- 102
\right] a^3
\nonumber \\
&&
+ \left[
- \frac{142793}{48}
- \frac{961}{54} \Nf^2
- \frac{80}{9} \zeta_3 \Nf^2
+ \frac{1427}{12} \zeta_3 \Nf
+ \frac{3663}{8} \zeta_3
+ \frac{39043}{72} \Nf
\right] a^4
\nonumber \\
&&
+ \left[
- \frac{242516239}{1296}
- \frac{60944155}{3888} \zeta_5 \Nf
- \frac{11155843}{1944} \zeta_3 \Nf
- \frac{2700601}{972} \Nf^2
\right. \nonumber \\
&& \left. ~~~~
- \frac{112}{9} \zeta_3 \Nf^3
+ \frac{3920}{81} \Nf^3
+ \frac{6800}{9} \zeta_5 \Nf^2
+ \frac{29501}{81} \zeta_3 \Nf^2
+ \frac{39000269}{1296} \zeta_3
\right. \nonumber \\
&& \left. ~~~~
+ \frac{41806447}{972} \Nf
+ \frac{137211305}{2592} \zeta_5
\right] a^5
\nonumber \\
&&
+ \left[
- \frac{251314718599}{20736}
- \frac{206090400131}{36864} \zeta_7
- \frac{78145127179}{62208} \zeta_3 \Nf
\right. \nonumber \\
&& \left. ~~~~
- \frac{27622282655}{20736} \zeta_5 \Nf
- \frac{10412288095}{6912} \zeta_3^2
+ \frac{2490660389}{10368} \zeta_3^2 \Nf
\right. \nonumber \\
&& \left. ~~~~
+ \frac{26119820537}{4608} \zeta_3
+ \frac{47880587209}{55296} \zeta_7 \Nf
+ \frac{107580934405}{31104} \Nf
\right. \nonumber \\
&& \left. ~~~~
+ \frac{351316606415}{41472} \zeta_5
- \frac{1235249951}{3888} \Nf^2
- \frac{4484669}{288} \zeta_7 \Nf^2
- \frac{987613}{72} \zeta_3^2 \Nf^2
\right. \nonumber \\
&& \left. ~~~~
- \frac{341953}{162} \zeta_3 \Nf^3
- \frac{3969}{4} \zeta_7 \Nf^3
- \frac{2885}{27} \Nf^4
- \frac{640}{27} \zeta_5 \Nf^4
+ \frac{464}{27} \zeta_3 \Nf^4
\right. \nonumber \\
&& \left. ~~~~
+ \frac{1465}{9} \zeta_5 \Nf^3
+ \frac{7676}{27} \zeta_3^2 \Nf^3
+ \frac{10511585}{972} \Nf^3
+ \frac{54873371}{648} \zeta_3 \Nf^2
\right. \nonumber \\
&& \left. ~~~~
+ \frac{75156965}{1296} \zeta_5 \Nf^2
\right] a^6 ~+~ O(a^7) ~.
\end{eqnarray}
The expressions for an arbitrary colour group have the same $\zeta_n$ 
dependence as is evident in the data file associated with the article.

\sect{Landau gauge $\MOMtccgzcs$ results.}

In order to make contact with previous results we provide the Landau gauge
$SU(3)$ anomalous dimensions for the $\MOMtccgzcs$ scheme as they can be 
compared directly with the five loop $\mMOM$ results of \cite{43,63,64}. This 
equivalence serves in part as a check on our symbolic manipulation code but 
also emphasises that the Landau gauge sector of the $\mMOM$ scheme involves 
neither $\zeta_4$ nor $\zeta_6$. In addition to the $\beta$-function of the
previous Appendix we have
\begin{eqnarray}
\gamma^{SU(3)}_{A,\MOMtccgzcss}(a,0) &=&
\left[
- \frac{13}{2}
+ \frac{2}{3} \Nf
\right] a
+ \left[
- \frac{255}{4}
+ \frac{67}{6} \Nf
\right] a^2
\nonumber \\
&&
+~ \left[
- \frac{8637}{4}
- \frac{719}{54} \Nf^2
- \frac{229}{12} \zeta_3 \Nf
- \frac{8}{9} \zeta_3 \Nf^2
+ \frac{11227}{24} \Nf
+ 324 \zeta_3
\right] a^3
\nonumber \\
&&
+~ \left[
- \frac{27189875}{256}
- \frac{1118977}{648} \Nf^2
- \frac{921265}{144} \zeta_5 \Nf
- \frac{889231}{144} \zeta_3 \Nf
\right. \nonumber \\
&& \left. ~~~~
- \frac{16}{9} \zeta_3 \Nf^3
+ \frac{665}{27} \Nf^3
+ \frac{5143}{27} \zeta_3 \Nf^2
+ \frac{9280}{27} \zeta_5 \Nf^2
+ \frac{1950705}{128} \zeta_5
\right. \nonumber \\
&& \left. ~~~~
+ \frac{5549393}{192} \Nf
+ \frac{7740879}{256} \zeta_3
\right] a^4
\nonumber \\
&&
+~ \left[
- \frac{71363464263}{16384} \zeta_7
- \frac{16520894997}{2048}
- \frac{16359945025}{6912} \zeta_5 \Nf
\right. \nonumber \\
&& \left. ~~~~
- \frac{8539017539}{20736} \zeta_3 \Nf
+ \frac{3150668061}{2048} \zeta_3
+ \frac{23633674547}{18432} \zeta_7 \Nf
\right. \nonumber \\
&& \left. ~~~~
+ \frac{29623505625}{4096} \zeta_5
+ \frac{327291152977}{124416} \Nf
- \frac{613783375}{2592} \Nf^2
\right. \nonumber \\
&& \left. ~~~~
- \frac{25237429}{432} \zeta_7 \Nf^2
- \frac{979887}{64} \zeta_3^2 \Nf
- \frac{292855}{54} \zeta_5 \Nf^3
\right. \nonumber \\
&& \left. ~~~~
- \frac{51647}{36} \zeta_3^2 \Nf^2
- \frac{1861}{27} \Nf^4
- \frac{304}{27} \zeta_3 \Nf^4
+ \frac{1760}{27} \zeta_5 \Nf^4
+ \frac{2240}{27} \zeta_3^2 \Nf^3
\right. \nonumber \\
&& \left. ~~~~
+ \frac{129545}{162} \zeta_3 \Nf^3
+ \frac{2420705}{324} \Nf^3
+ \frac{13148441}{2592} \zeta_3 \Nf^2
\right. \nonumber \\
&& \left. ~~~~
+ \frac{246191965}{1296} \zeta_5 \Nf^2
+ \frac{671260095}{2048} \zeta_3^2
\right] a^5 ~+~ O(a^6)
\nonumber \\
\gamma^{SU(3)}_{c,\MOMtccgzcss}(a,0) &=&
-~ \frac{9}{4} a
+ \left[
- \frac{153}{8}
+ \frac{3}{4} \Nf
\right] a^2
\nonumber \\
&&
+~ \left[
- \frac{11691}{16}
- \frac{5}{2} \Nf^2
+ \frac{9}{4} \zeta_3 \Nf
+ \frac{1269}{16} \zeta_3
+ \frac{1401}{16} \Nf
\right] a^3
\nonumber \\
&&
+~ \left[
- \frac{16985133}{512}
- \frac{16059}{32} \zeta_3 \Nf
- \frac{13115}{48} \Nf^2
- \frac{5895}{16} \zeta_5 \Nf
+ \frac{5}{2} \Nf^3
\right. \nonumber \\
&& \left. ~~~~
+ \frac{739053}{128} \Nf
+ \frac{1140075}{256} \zeta_5
+ \frac{2264193}{512} \zeta_3
+ 26 \zeta_3 \Nf^2
\right] a^4
\nonumber \\
&&
+~ \left[
- \frac{15114361941}{32768} \zeta_7
- \frac{9464865363}{4096}
+ \frac{9758887695}{8192} \zeta_5
\right. \nonumber \\
&& \left. ~~~~
- \frac{424982943}{4096} \zeta_3^2
- \frac{34433025}{256} \zeta_5 \Nf
- \frac{12230645}{288} \Nf^2
\right. \nonumber \\
&& \left. ~~~~
- \frac{86231}{2} \zeta_3 \Nf
- \frac{3969}{2} \zeta_7 \Nf^2
- \frac{219}{2} \zeta_3^2 \Nf^2
+ \frac{15353}{12} \Nf^3
\right. \nonumber \\
&& \left. ~~~~
+ \frac{477395}{192} \zeta_3 \Nf^2
+ \frac{550645}{96} \zeta_5 \Nf^2
+ \frac{814149}{128} \zeta_3^2 \Nf
+ \frac{34354859}{1024} \zeta_7 \Nf
\right. \nonumber \\
&& \left. ~~~~
+ \frac{1106092719}{4096} \zeta_3
+ \frac{1670942731}{3072} \Nf
- 65 \zeta_5 \Nf^3
- 14 \Nf^4
+ \zeta_3 \Nf^3
\right] a^5
\nonumber \\
&&
+~ O(a^6)
\end{eqnarray}
and
\begin{eqnarray}
\gamma^{SU(3)}_{\psi ,\MOMtccgzcss}(a,0) &=&
\left[
- \frac{4}{3} \Nf
+ \frac{67}{3}
\right] a^2
+ \left[
- \frac{706}{9} \Nf
- \frac{607}{2} \zeta_3
+ \frac{8}{9} \Nf^2
+ \frac{29675}{36}
+ 16 \zeta_3 \Nf
\right] a^3
\nonumber \\
&&
+~ \left[
- \frac{21683117}{648} \zeta_3
- \frac{2393555}{324} \Nf
- \frac{272}{9} \zeta_3 \Nf^2
- \frac{40}{9} \Nf^3
+ \frac{2861}{9} \Nf^2
\right. \nonumber \\
&& \left. ~~~~
+ \frac{74440}{27} \zeta_3 \Nf
+ \frac{15846715}{1296} \zeta_5
+ \frac{31003343}{648}
- 830 \zeta_5 \Nf
\right] a^4
\nonumber \\
&&
+~ \left[
- \frac{94958116621}{31104} \zeta_3
- \frac{26588447977}{27648} \zeta_7
+ \frac{2313514793}{10368} \zeta_3^2
\right. \nonumber \\
&& \left. ~~~~
+ \frac{14723323093}{5184}
+ \frac{18607183745}{7776} \zeta_5
- \frac{667846415}{1944} \zeta_5 \Nf
\right. \nonumber \\
&& \left. ~~~~
- \frac{251804567}{432} \Nf
- \frac{4726621}{243} \zeta_3 \Nf^2
- \frac{596849}{54} \zeta_3^2 \Nf
- \frac{80606}{81} \Nf^3
\right. \nonumber \\
&& \left. ~~~~
- \frac{6811}{2} \zeta_7 \Nf^2
- \frac{3520}{27} \zeta_5 \Nf^3
- \frac{128}{9} \zeta_3^2 \Nf^2
+ \frac{160}{27} \Nf^4
+ \frac{24064}{81} \zeta_3 \Nf^3
\right. \nonumber \\
&& \left. ~~~~
+ \frac{1063237}{27} \Nf^2
+ \frac{3085750}{243} \zeta_5 \Nf^2
+ \frac{4429579}{36} \zeta_7 \Nf
\right. \nonumber \\
&& \left. ~~~~
+ \frac{1748707267}{3888} \zeta_3 \Nf
\right] a^5 ~+~ O(a^6) ~. 
\end{eqnarray}
The quark mass dimension is
\begin{eqnarray}
\gamma^{SU(3)}_{m,\MOMtccgzcss}(a,0) &=&
-~ 4 a
+ \left[
- \frac{209}{3}
+ \frac{4}{3} \Nf
\right] a^2
\nonumber \\
&&
+~ \left[
- \frac{95383}{36}
- \frac{176}{9} \zeta_3 \Nf
- \frac{8}{3} \Nf^2
+ \frac{4742}{27} \Nf
+ \frac{5635}{6} \zeta_3
\right] a^3
\nonumber \\
&&
+~ \left[
- \frac{182707879}{1296}
- \frac{309295}{48} \zeta_5
- \frac{159817}{27} \zeta_3 \Nf
- \frac{13651}{27} \Nf^2
\right. \nonumber \\
&& \left. ~~~~
- \frac{3200}{9} \zeta_5 \Nf
+ \frac{8}{3} \Nf^3
+ \frac{1552}{9} \zeta_3 \Nf^2
+ \frac{5246557}{324} \Nf
+ \frac{15752321}{216} \zeta_3
\right] a^4
\nonumber \\
&&
+~ \left[
- \frac{75504232175}{7776}
+ \frac{3576071485}{27648} \zeta_7
+ \frac{9610932889}{5832} \Nf
\right. \nonumber \\
&& \left. ~~~~
+ \frac{17917034005}{31104} \zeta_5
+ \frac{187324052147}{31104} \zeta_3
- \frac{310328447}{432} \zeta_3^2
\right. \nonumber \\
&& \left. ~~~~
- \frac{257106335}{324} \zeta_3 \Nf
- \frac{180251015}{1944} \zeta_5 \Nf
- \frac{22459484}{243} \Nf^2
\right. \nonumber \\
&& \left. ~~~~
- \frac{4778536}{81} \zeta_7 \Nf
- \frac{60928}{81} \zeta_3^2 \Nf^2
- \frac{28096}{81} \zeta_3 \Nf^3
- \frac{1600}{9} \zeta_5 \Nf^3
\right. \nonumber \\
&& \left. ~~~~
- \frac{352}{27} \Nf^4
+ \frac{1372}{3} \zeta_7 \Nf^2
+ \frac{464038}{243} \Nf^3
+ \frac{948548}{27} \zeta_3 \Nf^2
\right. \nonumber \\
&& \left. ~~~~
+ \frac{1850845}{243} \zeta_5 \Nf^2
+ \frac{6570181}{162} \zeta_3^2 \Nf
\right] a^5 ~+~ O(a^6) ~.
\end{eqnarray}
It is straightforward to verify that these expressions tally with those in
\cite{43,63,64}.

\end{document}